\documentclass[a4paper,11pt]{article}
\pdfoutput=1 % if your are submitting a pdflatex (i.e. if you have
\usepackage{jinstpub} % for details on the use of the package, please
\usepackage[T5,T1]{fontenc} % From IceCube author list

\usepackage{lineno}
\usepackage{subfig}
\usepackage{tikz}
\usepackage{tikz-3dplot}
\tdplotsetmaincoords{60}{110}
\pgfmathsetmacro{\rvec}{.8}
\pgfmathsetmacro{\thetavec}{30}
\pgfmathsetmacro{\phivec}{60}

\usetikzlibrary{arrows,shapes,calc,trees,positioning,chains,shapes.geometric,%
    decorations.pathreplacing,decorations.pathmorphing, matrix,shapes.symbols}
\usetikzlibrary{positioning}
\tikzset{
    line/.style={draw, -stealth, very thick},
    mynode/.style={rectangle,rounded corners,draw=black, top color=white, bottom color=white,very thick, inner sep=1em, minimum size=3em, text width=14em, text centered},
    mynode1/.style={rectangle,rounded corners,draw=black, top color=white, bottom color=white,very thick, inner sep=0.2em, minimum size=3em,text width=0.25\columnwidth, text centered},
    mynode1aa/.style={rectangle,rounded corners,draw=black, top color=white, bottom color=white,very thick, inner sep=0.2em, minimum size=3em,text width=0.35\columnwidth, text centered},
    mynode1a/.style={rectangle,rounded corners,draw=YellowOrange!80, top color=white, bottom color=white,very thick, inner sep=0.2em, minimum size=3em,text width=0.35\columnwidth, text centered},
    mynode2/.style={rectangle,rounded corners,draw=white, top color=white, bottom color=white, very thick, inner sep=0em, minimum size=0em},
    mynode3/.style={rectangle,rounded corners,draw=YellowOrange!80, top color=white, bottom color=YellowOrange!15!white,very thick, inner sep=1em, minimum size=0em},
    mynode4/.style={rectangle,rounded corners,draw=black, top color=white, bottom color=black!15,very thick, inner sep=0.5em, minimum size=3em, text width=8em, text centered},
    myarrow/.style={->, >=latex', shorten >=2pt, thick},
    mylabel/.style={text width=7em, text centered},     
}  
\usetikzlibrary{calc}
\usepackage{xcolor}
\definecolor{newred}{RGB}{209,53,43}
\definecolor{newblue}{RGB}{74,125,179}

\usepackage{hyperref}

\usepackage{xspace}
\usepackage{amsmath}
\newcommand{\labfig}[1]{\label{fig:#1}}
\newcommand{\reffig}[1]{\hyperref[fig:#1]{Figure}~\ref{fig:#1}\xspace}

\newcommand{\labsec}[1]{\label{sec:#1}}
\newcommand{\refsec}[1]{\hyperref[sec:#1]{Section}~\ref{sec:#1}\xspace}

\newcommand{\labapp}[1]{\label{app:#1}}
\newcommand{\refapp}[1]{\hyperref[app:#1]{Appendix}~\ref{app:#1}\xspace}

\newcommand{\pdf}{PDF\xspace}
\newcommand{\pdfs}{PDFs\xspace}
\newcommand{\ssreco}{\emph{SegmentedSplineReco}\xspace}
\newcommand{\splinereco}{\emph{SplineReco}\xspace}

\usepackage{booktabs}
\usepackage{multirow}
\usepackage{array}

\hypersetup
{
    colorlinks=true,
    linkcolor=newred,
    urlcolor={newblue},
    filecolor={newblue},
    citecolor={newblue},
}

\title{A muon-track reconstruction exploiting stochastic losses for large-scale Cherenkov detectors}

\author[16]{R. Abbasi,}
\author[57]{M. Ackermann,}
\author[17]{J. Adams,}
\author[11]{J. A. Aguilar,}
\author[21]{M. Ahlers,}
\author[48]{M. Ahrens,}
\author[27]{C. Alispach,}
\author[30]{A. A. Alves Jr.,}
\author[40]{N. M. Amin,}
\author[13]{R. An,}
% ...
\affiliation[11]{Universit{\'e} Libre de Bruxelles, Science Faculty CP230, B-1050 Brussels, Belgium}
% ...
\affiliation[13]{Department of Physics and Laboratory for Particle Physics and Cosmology, Harvard University, Cambridge, MA 02138, USA}
% ...
\affiliation[16]{Department of Physics, Loyola University Chicago, Chicago, IL 60660, USA}
\affiliation[17]{Dept. of Physics and Astronomy, University of Canterbury, Private Bag 4800, Christchurch, New Zealand}
% ...
\affiliation[21]{Niels Bohr Institute, University of Copenhagen, DK-2100 Copenhagen, Denmark}
% ...
\affiliation[27]{D{\'e}partement de physique nucl{\'e}aire et corpusculaire, Universit{\'e} de Gen{\`e}ve, CH-1211 Gen{\`e}ve, Switzerland}
% ...
\affiliation[30]{Karlsruhe Institute of Technology, Institute for Astroparticle Physics, D-76021 Karlsruhe, Germany }
% ...
\affiliation[40]{Bartol Research Institute and Dept. of Physics and Astronomy, University of Delaware, Newark, DE 19716, USA}
% ...
\affiliation[48]{Oskar Klein Centre and Dept. of Physics, Stockholm University, SE-10691 Stockholm, Sweden}
% ...
\affiliation[57]{DESY, D-15738 Zeuthen, Germany}
% ...

\emailAdd{analysis@icecube.wisc.edu}
\abstract{IceCube is a cubic-kilometer Cherenkov telescope operating at the South Pole. The main goal of IceCube is the detection of astrophysical neutrinos and the identification of their sources.
High-energy muon neutrinos are observed via the secondary muons produced in charge current interactions with nuclei in the ice. Currently, the best performing muon track directional reconstruction is based on a maximum likelihood method using the arrival time distribution of Cherenkov photons registered by the experiment's photomultipliers. A known systematic shortcoming of the prevailing method is to assume a continuous energy loss along the muon track.  However at energies $>1$~TeV the light yield from muons is dominated by stochastic showers. This paper discusses a generalized ansatz where the expected arrival time distribution is parametrized by a stochastic muon energy loss pattern. This more realistic parametrization of the loss profile leads to an improvement of the muon angular resolution of up to 20\% for through-going tracks and up to a factor 2 for starting tracks over existing algorithms. Additionally, the procedure to estimate the directional reconstruction uncertainty has been improved to be more robust against numerical errors.}


\collaboration[c]{on behalf of the IceCube collaboration}

\toccontinuoustrue

\begin{document}
\maketitle
\flushbottom

\section{Introduction}
\labsec{intro}
The IceCube Neutrino Observatory is a cubic-kilometre neutrino telescope located at the geographic South Pole \cite{ABBASI2009294, ICinstrumentation}. 
It consists of 5160 digital optical modules (DOMs), each containing a 10-inch photomultiplier tube (PMT). The PMTs detect Cherenkov photons emitted from charged secondary particles created in neutrino interactions in the surrounding ice.
The ice in which IceCube is deployed is of glacial origin and exceptionally pure. However, it contains impurities such as dust and volcanic ash, most prominently in a layer between $\sim 2000$~m and $\sim 2100$~m depth  \cite{icecores, glaciology_2013, icemodel_paper}. Further irregularities in the ice include bubble columns in the refrozen ice around the strings of DOMs, a tilt of the ice sheet, and an anisotropic attenuation aligned with the local flow of the ice \cite{SpiceLea,Chirkin:2019vyq}. These medium qualities are partially mitigated by IceCube’s calibration system. A series of light-emitting diodes (LEDs) are used to illuminate the PMTs and parametrize the ice properties \cite{icemodel_paper}.

In 2013, the IceCube collaboration detected the first astrophysical neutrinos in the TeV-PeV range \cite{Aartsen:2014gkd}. Since then, further studies have been initiated to understand the origin of these neutrinos, e.g. \cite{galactic_constraints, blazar_contribution, anisotropies}. Many of these are point-source studies searching for correlations between an excess in neutrino events and known astrophysical source locations  \cite{Aartsen:2016oji}.
The coincidence in 2017 of a high energy neutrino event and the flaring blazar TXS~0506+056 \cite{txspaper} reinforced the idea of a fraction of blazars being the sources of high-energy neutrinos. In point-source analyses, a precise reconstruction of the direction of the neutrino is a central aspect that contributes most to the detection sensitivity. Apart from time integrated searches that collect large statistics samples, single real-time neutrino alerts that are sent out to the astronomical community also require a precise directional reconstruction \cite{ICalert_paper}.

An important detection channel for point source identifications are muons which originate from $\nu_\mu$ charged current (CC) interactions, which appear as track signatures in the IceCube detector. Above a few TeV, these muons are nearly co-aligned with the direction of the parent neutrino due to relativistic kinematics (see \refsec{oldrecos} for a more quantitative discussion). At these energies, a confidence interval on the muon arrival direction approximately translates into a confidence interval on the parent neutrino direction.

This paper describes a new likelihood reconstruction that incorporates the stochastic energy losses of muons along their tracks into the likelihood hypothesis and thereby improves the accuracy and precision of the reconstructed arrival direction. \refsec{oldrecos} summarizes the previous likelihood approaches and lists their limitations. \refsec{ssreco} introduces the new likelihood approach and discusses how some shortcomings of the previous algorithms are solved. The discussion here concentrates around both the angular reconstruction and uncertainty estimation. \refsec{results} and \refsec{disc} show comparisons of the various methods and conclude this manuscript with a final discussion.

\section{Previous algorithms}
\labsec{oldrecos}
This section illustrates a typical reconstruction process and in this context describes various existing reconstructions that are useful to understand the benefits of the new approach. Some of the details in this section have been already partially covered in Refs. \cite{amanda_reco} and \cite{paraboloid}.

\subsection{Angular reconstruction}
IceCube collects Cherenkov photons from charged particles with a few nanosecond time resolution at the various DOM locations \cite{ICinstrumentation}. This time resolution is required because a Cherenkov light front from a muon passing by a DOM yields a photon arrival time probability distribution function (\pdf) that can have temporal structures of this magnitude \cite{light_propagation_wiebusch_raedl}. This time resolution can best be exploited in an unbinned likelihood approach. 
The prevailing angular reconstructions typically assume an infinite muon track length and neglect stochastic losses. Such reconstructions employ two broad classes of unbinned likelihood approaches to model the arrival time of photons:
\begin{align}
L(x,y,z,t,\theta, \phi)&=\prod\limits_{j=1}^{N_{\mathrm{DOM}}}\prod\limits_{i=1}^{N_{\mathrm{hit}}} [p_j(t_i)]^{q_i} \label{eq:llh_standard} \\
L_{\mathrm{1st}}(x,y,z,t,\theta, \phi)&=\prod\limits_{j=1}^{N_{DOM}} p_{j,\mathrm{1st}}(t_1) \nonumber \\
&\propto \prod\limits_{j=1}^{N_{DOM}} [p_j(t_1)]^{q_1} \cdot (1-P_j(t_1))^{Q_j-q_1}
\label{eq:llh_first}
\end{align}
where $N_{\mathrm{DOM}}$ and $N_{\mathrm{hit}}$ are the total number of DOMs and hits, respectively. The first likelihood (\autoref{eq:llh_standard}) is the \textit{standard unbinned likelihood} and uses the photon arrival \pdf $p(t)$ for each observed hit.
Since multiple photons can arrive at the DOM simultaneously, the correct application weights each hit by the observed charge $q_i$.
In practice, this is often neglected since the effect it has is subdominant compared to the unphysical assumption of the minimally ionizing track hypothesis.
The second likelihood (\autoref{eq:llh_first}) uses the photon arrival \pdf of only the first photon, $p_{j,\mathrm{1st}}(t_1)$, which technically corresponds to the  first-order-statistic PDF.\footnote{In a previous publication \cite{amanda_reco} the likelihood is called ``MPE" likelihood instead of $L_{\mathrm{1st}}$.} This PDF can be calculated exactly given the \pdf $p(t)$ and the cumulative distribution function (CDF) $P(t)$ \cite{amanda_reco}. The likelihood uses data that is weighted with the observed charge per hit $q_i$ {\cite{ICinstrumentation}}, and the total observed charged per DOM, $Q=\sum_i q_i$.
The motivation for the first-order statistic \pdf is twofold. While the standard \pdf (\autoref{eq:llh_standard}) is able to model the time distribution of a minimally ionizing muon in homogeneous ice, muons at energies beyond TeV energies undergo stochastic energy losses. Additionally, the real ice properties are not homogeneous. These known unphysical assumptions are partly mitigated by using $L_{\mathrm{1st}}$, because it evaluates only the first observed photon in each DOM which is less likely to have undergone significant scattering and hence is less affected by these uncertainties.
The optimization is a 6-dimensional problem for the track positional parameters $x$, $y$, $z$, $t$ and orientation via two angles, e.g. zenith $\theta$ and azimuth $\phi$. In practice, the angles are sometimes re-parametrized as two parameters lying in the sphere tangent plane that is perpendicular to the seed direction in order to avoid numerical issues near the poles (see \refapp{appendix_angle_parametrizations}).

%% SEED CHAIN Graph: 
%% linefit->analytic fit -> B-Spline fit -> Seg Spline Reco
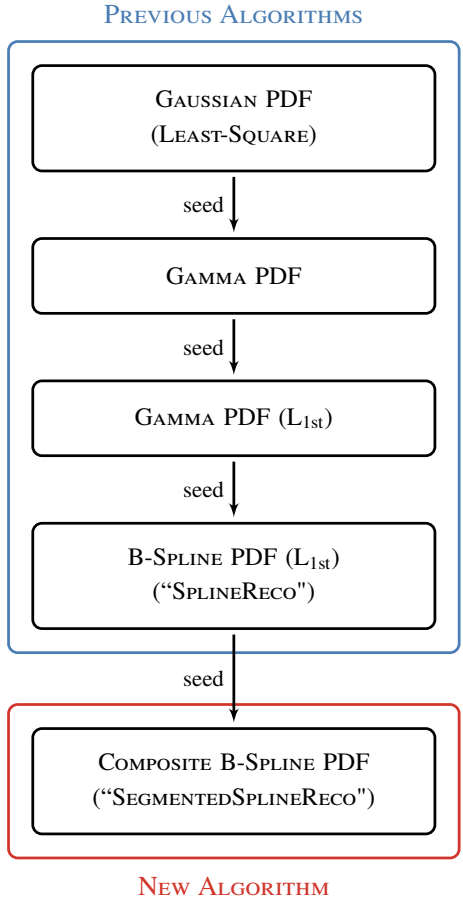
\begin{figure}[t]
\begin{center}
% \resizebox{0.55\columnwidth}{!}{%
\resizebox{0.4\columnwidth}{!}{%
\begin{tikzpicture}[node distance=0cm, auto, font = \normalsize]  
\node[mynode] (linefit) { \textsc{Gaussian \pdf \\ (Least-Square)}};
\node[mynode, below=1 cm of linefit] (analyticSPE) { \textsc{Gamma PDF}};
\node[mynode, below=1 cm of analyticSPE] (analyticMPE) { \textsc{Gamma \pdf ($\mathrm{L}_{\mathrm{1st}}$)}};
?\node[mynode, below=1 cm of analyticMPE] (spline) { \textsc{B-Spline \pdf ($\mathrm{L}_{\mathrm{1st}}$) \\ (``SplineReco")}};
?\node[mynode, below=1.5 cm of spline] (ssreco) { \textsc{Composite B-Spline \pdf \\ (``SegmentedSplineReco")}};

%Larger rectangles
\draw[newblue, very thick, rectangle, rounded corners] ($(linefit.north west)+(-0.35,0.35)$)  rectangle ($(spline.south east)+(0.35,-0.35)$);
\draw[newred, very thick, rectangle, rounded corners] ($(ssreco.north west)+(-0.35,0.35)$)  rectangle ($(ssreco.south east)+(0.35,-0.35)$);

\node[ text centered, above=0.5 cm of linefit, newblue] (oldreco) {\large \textsc{Previous Algorithms}};
\node[ text centered, below=0.5 cm of ssreco, newred] (newreco) {\large \textsc{New Algorithm}};

% \node (rect) [rectangle, rounded corners, draw, minimum width=0cm, minimum height=5cm, black, thick, above= 0.4cm of linefit] {} node [fill=white, midway, below, yshift=+1.7cm] {\small \color[newblue]{\textsc{Old recos}}};

\draw[->, >=latex', shorten >=2pt, shorten <=2pt, bend right=0, very thick, solid, color=black] (linefit.south) to node[auto, swap] {\small seed}(analyticSPE.north); 
\draw[->, >=latex', shorten >=2pt, shorten <=2pt, bend right=0,  very thick, solid, color=black] (analyticSPE.south) to node[auto, swap] {\small seed}(analyticMPE.north); 
\draw[->, >=latex', shorten >=2pt, shorten <=2pt, bend right=0,  very thick, solid, color=black] (analyticMPE.south) to node[auto, swap] {\small seed}(spline.north); 
\draw[->, >=latex', shorten >=2pt, shorten <=2pt, bend right=0,  very thick, solid, color=black] (spline.south) to node[auto, swap] {\small seed}(ssreco.north); 
\end{tikzpicture} 
}%
\end{center}
\caption[]{\labfig{reco_chain}An example track reconstruction seeding chain depicting the underlying assumed PDF in the respective likelihood reconstruction. The \pdf approximation quality and {CPU} time requirements broadly increase with more sophisticated reconstructions later in the chain.}
\end{figure}

The following sections describe a processing pipeline in which several reconstructions with increasing complexity are applied to the same event (see \reffig{reco_chain}). Each reconstruction outcome is used as a seeding strategy to the next one, starting with the fastest and simplest reconstruction and becoming gradually more time consuming and more precise. Note that in practice different processing pipelines are used based on the needs of a given event selection.
However, the pipeline illustrated in \reffig{reco_chain} is representative in that it shares the property of starting simple and fast and becoming gradually more time consuming and precise.\footnote{There is an exception in the time consumption of $L_{1st}$, which only requires the evaluation of the first photon, and is therefore actually faster, while also giving a more precise result.} Furthermore, it illustrates the thought process that went into the development of the new reconstruction.

\paragraph{Analytic Gaussian \pdf (Least-Square) with plane-wave assumption}
The ``first-guess" fit is typically a least-square fit which does not assume a minimal-ionizing muon, but a plane-wave moving through the detector with a constant velocity.\footnote{This algorithm has been previously been referred to as ``line-fit"\cite{amanda_reco}.} Assuming a standard normal arrival time \pdf of this plane wave one can analytically calculate a mean position and velocity vector of the least-square problem \cite{original_linefit}. To partially mitigate this rather simplified approximation a more robust regression with respect to outliers is typically used, for example by the inclusion of a Huber loss term \cite{improved_linefit}. The robust algorithm is shown in \reffig{linefit_pandel}. 

\paragraph{Analytic Gamma PDF}
% PANDEL
A more physically motivated likelihood description of the photon arrival \pdf comes from assuming a minimally ionizing track as discussed above. This can be analytically described with a gamma distribution \cite{amanda_reco} and implicitly contains the Cherenkov emission geometry and scattering properties of homogeneous ice.\footnote{In a previous publication \cite{amanda_reco} this approximation of the PDF with a gamma function has been called ``Pandel" approximation.} This parametrization is more accurate not only because of its more realistic assumptions, but also because it allows to calculate analytically the first-order-statistic PDF and thereby $L_{\mathrm{1st}}$.
Empirically, the gamma PDF with likelihood $L_{\mathrm{1st}}$ gives better results than the Gaussian first guess (least-square fit) or the standard likelihood $L$, as shown in \reffig{linefit_pandel}. As described earlier, this can be understood as partly mitigating the inaccurate modeling assumption of an infinite muon by only looking at the first photon. Also {shown} is the mean angle between the muon and parent neutrino direction. The uncertainty from the displayed muon reconstructions is well above this intrinsic uncertainty, especially above 10-100~TeV.

\begin{figure}[t]
    \centering
    \includegraphics[width=0.7\columnwidth]{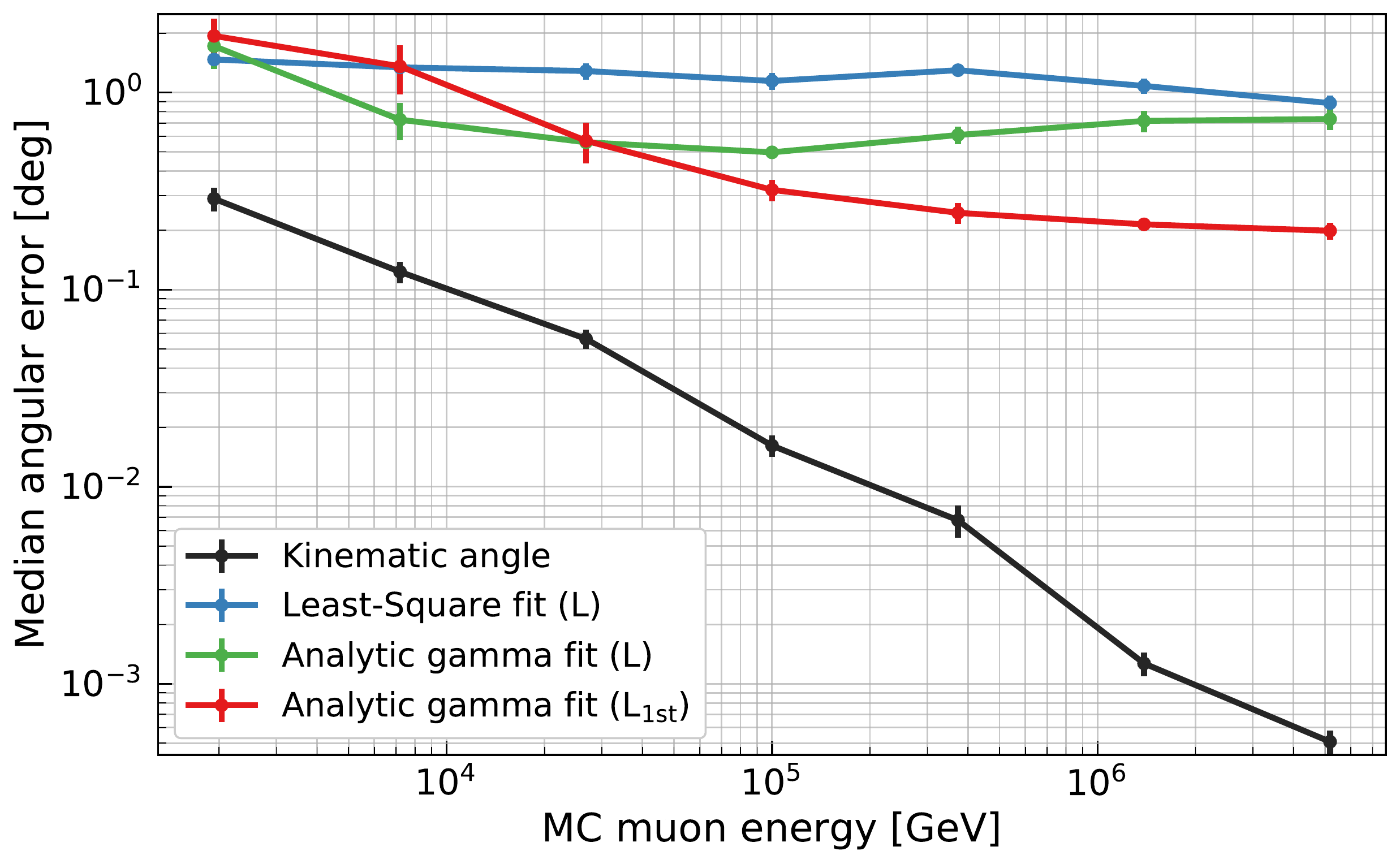}
    \caption{Median angular error for muons passing the whole detector (through-going muons) as a function of the simulated muon energy at the interaction vertex for a least-square fit and two different analytic Gamma fits. The kinematic angle between the neutrino and the muon direction is shown in black. Each reconstruction is seeded with the previous algorithm, following the chain of \reffig{reco_chain}. The statistical error on the median is calculated using bootstrapping.}
    \labfig{linefit_pandel}
\end{figure}

\begin{figure}
    \centering
    \includegraphics[width=0.7\columnwidth]{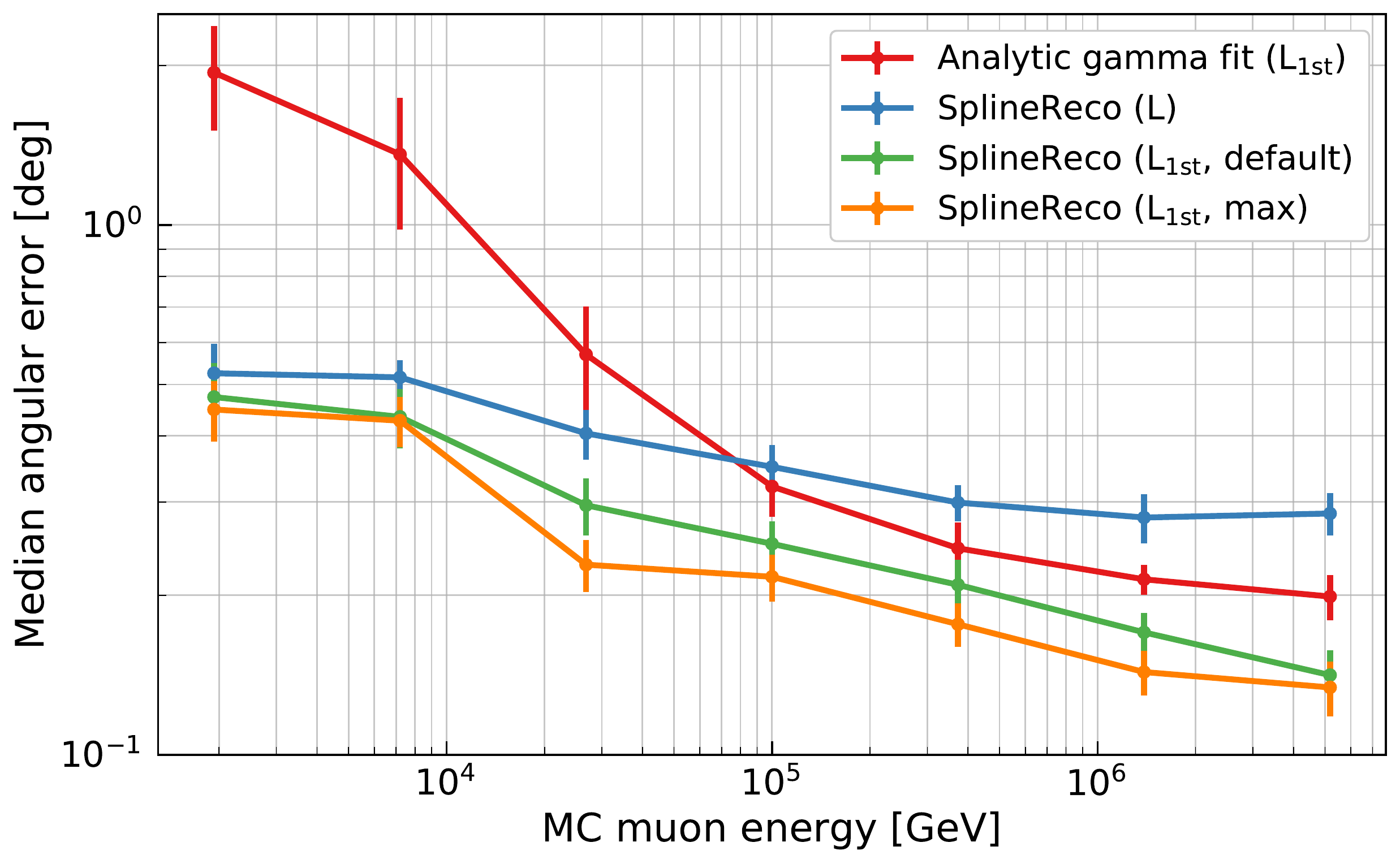}
    \caption{Reconstruction of {muons passing the whole detector (through-going muons)} using the analytic Gamma fit with the pdf$_{\mathrm{1st}}$ \pdfs and the \splinereco algorithms. The \splinereco reconstruction is shown using the standard likelihood $L$, $L_{\mathrm{1st}}$ with default settings, and $L_{\mathrm{1st}}$ with max settings. Each reconstruction is seeded with the previous best performing algorithm, following the chain of \reffig{reco_chain}. The statistical error on the median is calculated using bootstrapping. {The MC muon energy is calculated at the neutrino interaction point.}}
    \labfig{pandel_splinereco}
\end{figure}

\paragraph{B-Spline \pdf modelling}
Further improvement can be made {by incorporating a more realistic ice model}{, which involves two steps.} First, the ice's optical properties are inferred from fitting an ice model to LED flasher data \cite{icemodel_paper}. {Second}, infinite muon minimal-ionizing muon tracks are simulated in many different positions and orientations, and the resulting photons are recorded in high-dimensional histograms \cite{photonics_paper}. These histograms are then fitted with multi-dimensional interpolating B-Splines \cite{splines}  to represent the photon arrival \pdf in dependence of the track position and orientation. {Since the relevant ice properties match reality better, the dominant remaining inaccuracy comes from the fact that the stochastic losses of the muon are neglected, as the track hypothesis still assumes a minimally ionizing particle.} 
% In order to overcome this, several modifications have been introduced on top of the first-order statistic \pdf (usage of $L_{\mathrm{1st}}$ instead of $L$) which have been empirically observed to yield further {improvements}. 
{Several methods have been implemented to mitigate the effects of this unphysical assumption and improve the reconstruction, besides the use of the first-order statistic \pdf (usage of $L_{\mathrm{1st}}$ instead of $L$).}
% These include the use of effective photon arrival \pdfs from averaged stochastic tracks instead of minimally ionizing tracks, non-uniform photo-multiplier noise modeling, a removal of photons that might arise from large stochastic losses and an energy-dependent convolution of the first-order statistic PDF with a Gaussian kernel \cite{schatto}. 
{These include using effective photon arrival \pdfs from averaged stochastic tracks instead of minimally ionizing tracks, including non-uniform photo-multiplier noise modeling, removing photons that might arise from large stochastic losses, and convolving  the first-order statistic PDF with an energy-dependent Gaussian kernel \cite{schatto}.}
{We call this ensemble of modifications} ``max settings” in the following. %the use of these modifications  \emph{max settings} in the following.
The general reconstruction scheme with B-splines as described above {is referred to as} \splinereco. When no added information is given, \splinereco uses likelihood $L_{\mathrm{1st}}$ and the previously described  \emph{max settings}.
As can be seen in \reffig{pandel_splinereco}, usage of likelihood $L_{\mathrm{1st}}$ is better than the standard likelihood ($L$) {and the gamma distribution-based approach} (\reffig{linefit_pandel}). The use of the modifications we refer to as \emph{max settings} gives some additional improvement. Compared to the analytic gamma approximation, the \pdfs based on B-splines are {significantly} more precise and still reasonably fast. A typical application of the reconstruction on a 10~TeV muon takes less than a hundredth of a second. A more thorough comparison of running times is given in \refsec{runtime}. While the modifications {have been empirically shown to} somewhat circumvent the underlying unphysical assumption of an infinite muon track with a smooth energy loss profile, it is preferable to model a more correct hypothesis. The new algorithm that models stochastic losses directly is described in \refsec{ssreco}.

\subsection{Uncertainty estimation}
\labsec{paraboloid}

The previously discussed likelihood optimizations typically run over 6 parameters: $x$, $y$, $z$, $t$, $\theta$ and $\phi$. {The first four define a point lying on the track ($x$, $y$, $z$, $t$) and the last two specify the direction using zenith and azimuth ($\theta$, $\phi$).}
The angle parameters are often re-parametrized for technical reasons (see  \refapp{appendix_angle_parametrizations} for details). In order to obtain an uncertainty estimate for the two angles, the typical practice is to fit a 2-D paraboloid to the profile-likelihood at the minimum \cite{paraboloid}, where the $x$, $y$, $z$ and $t$ parameters are profiled, i.e. optimized for each value of the two angles. This uncertainty estimation is referred to as the {\emph{traditional method}} in the following. In practice, a grid is constructed in a rotated coordinate system $\varphi_1, \varphi_2$, which is localized at the equator such that the two new angle coordinates are comparable \cite{paraboloid} (see \refapp{app_equator_parametrization} for more information). A problem with this construction is that {the fixed grid used for the evaluation of the likelihood space is optimized} for a typical uncertainty of about a degree. If the actual uncertainty is much smaller or larger than this typical uncertainty, this can {lead} to failures on the paraboloid fit. Additionally, if a fit converges, it is not clear how well it describes the shape of the log-likelihood maximum. Especially at lower energies that shape is typically not parabolic.  In \refsec{ssreco} an updated strategy is described that avoids both the problem with {the fixed grid} and the fit quality check.

\section{New algorithm: SegmentedSplineReco}
\labsec{ssreco}

At energies above {$\sim1$~TeV,} muons predominantly {lose} energy stochastically via bremsstrahlung, pair production, and nuclear interactions. {Therefore,} the assumption of a continuous energy loss pattern used in the reconstructions presented up to now is no longer valid.
% The muon track reconstruction algorithms described in \refsec{oldrecos} assume that the muon is loosing energy continuously along its path. However, starting above $\sim 1$~TeV, muons predominantly lose energy stochastically via bremsstrahlung, pair production, and nuclear interactions. 
% The effect of these processes is the production of clustered light depositions on top of the track signature. 
The result of these stochastic energy losses is the production of clustered light depositions on top of the track signature. 
Light created in such stochastic losses has a different emission spectrum, and it influences the photon arrival time distribution. Therefore, these stochastic energy losses should ideally be included directly in the track parametrization. A new reconstruction implementing this idea is described in this section. This reconstruction is referred to as \ssreco in the following.

\begin{figure}[t]
    \centering
    \includegraphics[width=0.4\columnwidth]{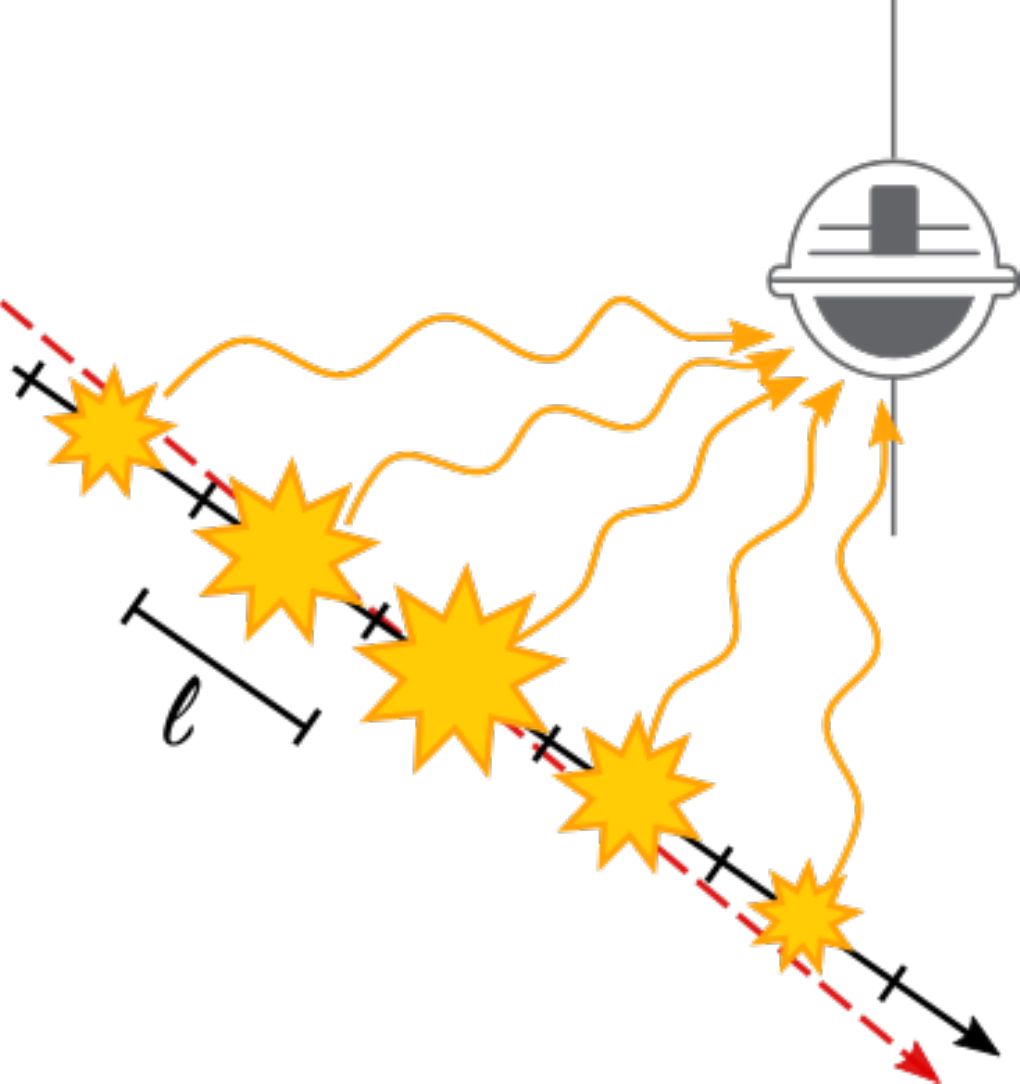}
    \caption{Schematic view of \ssreco. The incoming muon track in red is first reconstructed by the black line, representing the initial track hypothesis for \ssreco. It follows an energy reconstruction which results in a series of cascades along the muon track (yellow stars), placed at the center of each segment of length $\ell$. The energy information of each segment is used to define the final \pdf at each DOM using \autoref{eq:ssreco_pdf}.}
    \labfig{ssreco}
\end{figure}
% SplineMPE / SegmentedSplineMPE
% Motivation -> figure

% \begin{enumerate}
% \item SPE/MPE/ExtSPE likelihood
% \item New parametrization (Equator parametrization - from UberMillipede)
% \item Full gradient/Hessian calculation (in $x$, $y$, $z$, $\theta$, $\phi$, $t$, energies; SPE/MPE/ExtSPE)  
% \item Full utilization of the power of B-splines
% \item Energies as potential optimization pars
% \item Cascade energy fit
% \item Uncertainty estimation included
% \item Allow TimeWindow Cleaning of saturated pulses (non-trivial due to hessian)
% \item Possibility to reconstruct Prepulses
% \item Post-Jitter convolution
% \end{enumerate}

% \subsection{Stochastic energy losses}
% \subsection{Likelihood definitions}

\subsection{Angular reconstruction}
\labsec{ssreco_angres}

\ssreco is a maximum likelihood reconstruction that uses a segmented muon hypothesis (\reffig{ssreco}). Each segment effectively models electromagnetic and hadronic stochastic losses (``cascades"), and contributes to the \pdf of the photon arrival times, together with a constant DOM-dependent noise term and an optional infinite minimum ionizing muon track hypothesis. 
% {The same high-dimensional splines that are used for \textit{SplineReco} are used to obtain the number of photons and their time arrival time distributions.}
The number of photons and their time arrival distributions are obtained from high-dimensional splines fitted to Monte Carlo simulations of photons propagating in ice. {Similar splines for the infinite muon hypothesis are used in the \splinereco reconstruction. The main difference are the additional B-Splines for the stochastic losses in \ssreco.} %, while \splinereco only uses a single spline function for an infinite track. 

The reconstruction performs several steps which are described {below}:

\begin{enumerate}
   
   % \item The \emph{SplineMPE} reconstruction is performed giving a track as output.
   % \item The \emph{SplineMPE} track is used as a seed for an additional energy loss pattern reconstruction called \textbf{\emph{Millipede} described in \cite{enreco}}. The outcome of this reconstruction is a series of $n$ cascades of energy $E_j$ representing the stochastic energy losses fixed along the muon track.
    \item The initial hypothesis is twofold: (1) a track {direction} and (2) an energy loss pattern parametrized by electromagnetic cascades placed at the center of each segment and located along the initial track hypothesis (see \reffig{ssreco}). These first guesses are given by previous reconstructions. 
    Alternatively {the} energy loss pattern can also be determined directly by \ssreco, in which case only an initial guess of the track direction is required. 
    \item The total \pdf of the photon arrival time $t$ at a DOM position in the detector is given by the weighted sum of $n$ \pdfs using each cascade segmented as a source of photon emission:
    \begin{equation}
        \label{eq:ssreco_pdf}
        p(t) = \sum_{k=0}^{n} w_{k}p_{k}(t),
    \end{equation}
    where the index {$k$} runs over the $n$ segments and $w_{j} = \frac{\lambda_j}{\sum_{k=1}^n \lambda_k}$. The parameter $\lambda_j$ denotes the expected number of photons of the source in segment $j$ in the given DOM, where the different sources are the electromagnetic cascades, the constant noise contribution, and {the contribution of a minimum ionizing muon if requested.} The total number of photons from the cascades and the muon are again obtained from high-dimensional spline distributions fitted to simulations. \reffig{ssreco_pdf} shows the \pdf for all cascades produced along a muon track as a function of hit time, their weighted sum and the infinite muon and noise \pdfs.
    
    \begin{figure}
    \centering
    \includegraphics[width=0.7\columnwidth]{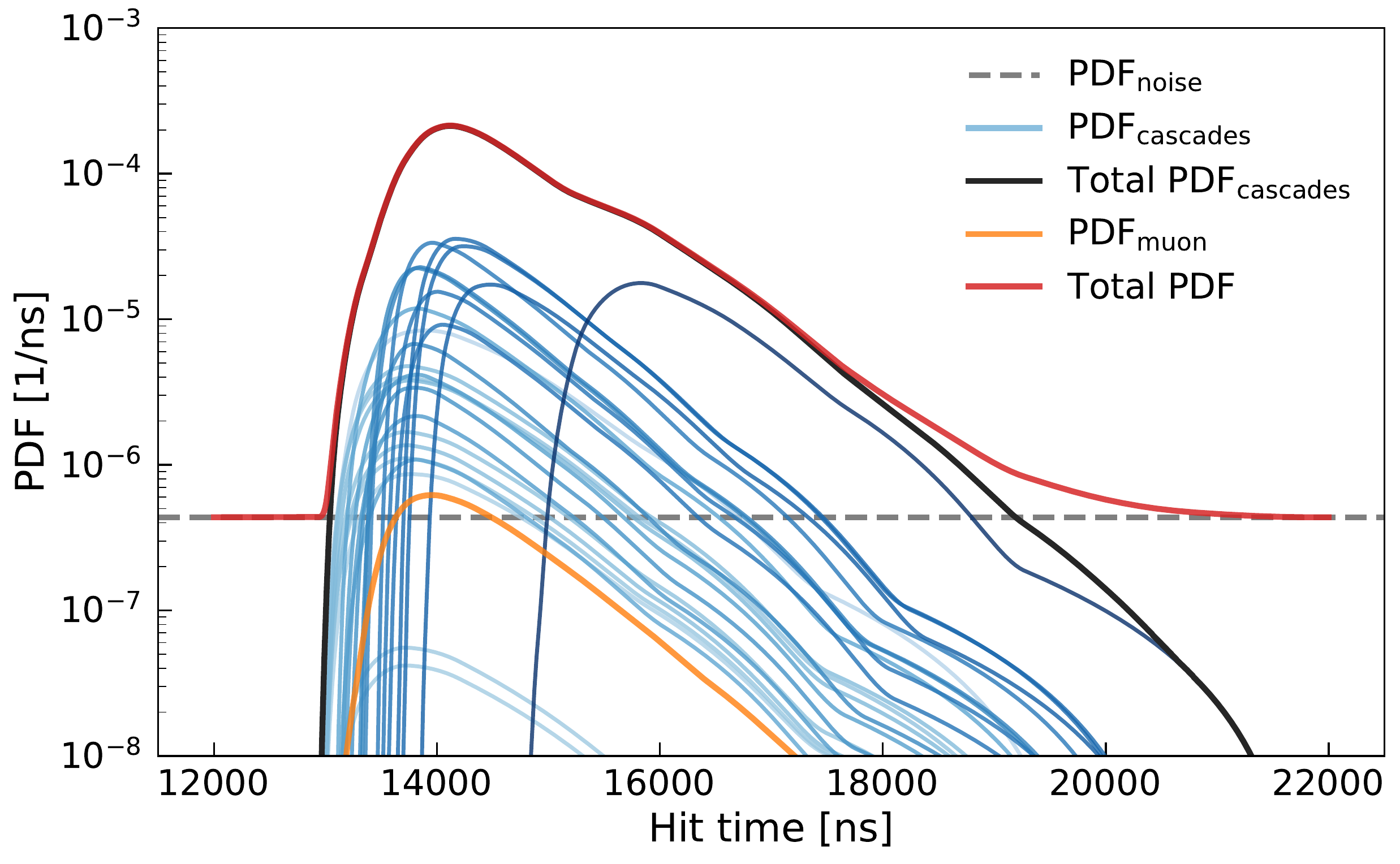}
    \caption{SegmentedSplineReco \pdf for one DOM as a function of the photons arrival time, according to \autoref{eq:ssreco_pdf}. The blue lines are the \pdfs of each cascade along the muon track in order of creation from lighter to darker blue; their weighted sum is given by the solid black line. The orange line shows the \pdf for a minimum ionizing muon, the dashed gray line the noise PDF. The total \pdf is shown by the red curve.}

    \labfig{ssreco_pdf}
    \end{figure}

    \item The \pdf is then used to define a likelihood function, which is maximized varying the track parameters ($x$, $y$, $z$, $t$, $\theta$, $\phi$).\footnote{Here we use the parameters $\theta$ and $\phi$, the usual angle parameters in spherical coordinates. In a more practical optimization these variables are re-parametrized in one of two different ways and then denoted either as $\varphi_1$ and $\varphi_2$ or $\Phi_1$ and $\Phi_2$ (see \refapp{appendix_angle_parametrizations}).}
    Three likelihood functions have been implemented:
    \begin{enumerate}
        \item standard unbinned likelihood 
        \begin{equation}
             , \mathrm{L}=\prod_j^{N_{\mathrm{DOM}}} \prod_i^{N_{\mathrm{hit}}} [p_j(t_i)]^{q_i},
        \end{equation}
        \item extended unbinned likelihood
        \begin{equation}
            \displaystyle \mathrm{L}_{\mathrm{ext}}=\prod_j^{N_{\mathrm{DOM}}} \frac{e^{-\lambda_j} {\lambda_j}^{q_j}}{q_j!} \prod_i^{N_{\mathrm{hit}}} [p_j(t_i)]^{q_i},
        \end{equation}
        \item unbinned likelihood for the first hit on each DOM: 
        \begin{equation}
            \displaystyle \mathrm{L}_{\mathrm{1st}}=\prod_j^{N_{\mathrm{DOM}}} p_{j,1\mathrm{st}}(t_1).
        \end{equation}
    \end{enumerate}
    The index $j$ runs over all DOMs while the index $i$ runs over all hits for a given DOM $j$. The total {charge} produced by a hit $i$ is denoted by $q_i$.
    The \pdf $p_{j,1}$ in likelihood (c) is the first-order-statistic PDF, this time derived from the total PDF of all source contributions. The derivation is mathematically similar to the one for a single minimal-ionizing muon $L_{\mathrm{1st}}$ in \autoref{eq:llh_first}.

%MPE (Multi Photo Electron), SPE (Single Photo Electron) and Extended-SPE. The MPE likelihood takes into account all light arriving at a DOM, but is adjusted to only use the time information of the first photon since less scattered. In SPE, the \pdf of equation \ref{eq:pdf} is calculated for each photon produced in the DOM, therefore all light information is used. Extended-SPE is obtained multiplying the SPE likelihood by a Poisson factor which takes into account the energy content of the event.

\end{enumerate}

\ssreco has been implemented in C++ and Python within the IceCube software framework. It includes several improvements with respect to the previous algorithms. This includes support for exact gradient {and calculation of the second-order partial derivatives matrix, the Hessian matrix,} from the underlying high-dimensional B-splines and the possibility to fit the energies jointly with the track parameters. The latter option is only feasible with available gradient information due to the rather high-dimensional ({$> 100$-D}) problem. Supplying the optimization algorithm with an exact gradient leads to substantially improved convergence speed. For the often used algorithm \emph{MIGRAD} contained within the high-energy physics package \emph{minuit} \cite{minuit} {the resulting increase in speed is} {a factor of two compared to not using a gradient}.

A new coordinate system has also been implemented, equivalent to the one described in \refsec{paraboloid}: the seed track is rotated to the equator of the coordinate system, defining a new origin. {The coordinates near the equator are quasi-euclidean for small values and remain interpretable as angles for larger values.} More details can be found in \refapp{appendix_angle_parametrizations}.
%Different minimization algorithm are available, among which Migrad, Minuit, Simplex and all \texttt{scipy} minimization options. 

% In \ssreco there is also the possibility to include the effect of prepulses in the \pdf and to perform the cleaning of saturated pulses, particularly relevant for the reconstruction of real time events.

Finally, an energy-dependent convolution of the first-order statistic {\pdf has} been implemented similar to the implementation in {\splinereco. The} \ssreco likelihood is convoluted with a Gaussian distribution using a fast recursive approximation algorithm as implemented in \cite{getreuer2013survey}. Such a convolution implicitly models timing inaccuracies between modules and also mitigates remaining model mis-specification. The likelihood is calculated at several sample points around the requested time. From these sample points, the convoluted \pdf is calculated with adjustable accuracy depending on sample point density and recursion step count. The convolution is sometimes called \emph{post-jitter} {below}. 
% The values of $\sigma_{\mathrm{LLH}}$ depend on the muon energy which is given by an energy estimator; they are listed in Table 4.1. The effect of the energy dependent post-jitter convolution on the angular resolution of SegmentedSplineReco using the MuonBenchmark datasets is shown in Figure 4.19. The results for SegmentedSplineReco using the MPE likelihood are compared to two SplineMPE reconstructions: the default one does not include the energy dependent post-jitter, on the other hand the max setting contain

\subsection{Uncertainty estimation}
\labsec{subsec:ssreco_error}
Two new uncertainty estimation methods have been implemented for \ssreco. Both of them estimate {the Hessian matrix} at the log-likelihood optimum.
The first approach (\emph{Method 1}) calculates it analytically using the ability of the B-spline \pdfs to yield exact higher-order derivatives.
The second approach (\emph{Method 2}) samples the 6-D minimum of the negative log-likelihood function in the track parameters ($x$, $y$, $z$, $t$, $\varphi_1$, $\varphi_2$)\footnote{The angles here are defined in the rotated parametrization (see \refapp{appendix_angle_parametrizations}).} with a Markov Chain Monte Carlo (MCMC) sampler. The result is used to fit a 6-D elliptic paraboloid to the {log-likelihood} landscape which again {requires the calculation of the $6\times6$ Hessian matrix.} Details are given in \refapp{uncertainty_details}. While this approach is more time consuming, it can be a little more robust against non-Gaussianities close to the optimum if a modified $\chi^2$ loss function is used (see \refapp{uncertainty_details}) and often leads to slightly wider contours as shown in \refsec{results}.

The inverse of the obtained Hessian yields the covariance matrix containing the parameter correlations. A reduction to the $2\times2$ submatrix of the angle parameters marginalizes the other parameters and results in a 2-D uncertainty ellipse for the direction. In comparison, the \emph{traditional method} described in \refsec{paraboloid} performs a paraboloid fit using profile likelihood evaluations in the two angular dimensions. {Performing these profile likelihood evaluations is time consuming} {and can lead to unstable {results,} while the fixed grid size makes it unadaptable to different uncertainty angular scales.} %, can lead to unstable results, and the fixed grid size makes it unflexible for different scales. This extra optimization is time consuming, can lead to unstable results, and the fixed grid size makes it unflexible for different scales.
{Both of these problems are solved with either of the new approaches.} {In particular, the new methods can detect when the outcome of the uncertainty estimation is unreliable as explained in \refapp{uncertainty_details}.}
% In particular, the new methods allow {detection of} when the uncertainty estimation is unreliable, as outlined in \refapp{uncertainty_details}.

Another possibility that comes with the analytic Hessian (\emph{Method 1}) is to jointly calculate the Hessian with respect to the six track parameters and additionally the energy parameters of all individual energy losses. Computationally, this \emph{full} calculation has nearly no overhead, but it broadens the final uncertainty contours over the two angular dimensions due to the extra marginalization over the energy dimensions if it is used. {This can be desirable, since the uncertainty is by construction too small if the energies of stochastic losses are fixed (see \refsec{sec:angular_error_estimation_new}). However, if the energies are fixed in the previous optimization procedure, and also due to the high dimensionality of the problem, the enlarged {Hessian} is usually not calculated at a local minimum and often not positive definite, so we do not use this procedure in practice.}
% In particular, we can decide to calculate the full covariance matrix  with respect to the energy parameters as well. The computational load of this full Hessian matrix has nearly no overhead but it broadens the final uncertainty contours due to the extra marginalization over the energy dimensions.

\section{Performance comparisons}
\labsec{results}
The new reconstruction has been applied {to three different selections of simulated muon track events.} The first dataset contains muon tracks that pass quality cuts (so-called NDir/LDir cuts\footnote{The exact cuts are $\mathrm{LDir} \geq 600$ and $\mathrm{NDir} \geq 8$.} on the number of hit DOMs and track length, respectively; see \cite{Aartsen:2016oji} for more information) based on \splinereco, which to some extent mimics events that are usually found on the final analysis selections used in IceCube. In the following they are referred to as \emph{SplineReco-optimized}. Events with large stochastic losses typically obtain low NDir/LDir values with \splinereco. These events subsequently do not pass the cuts and should be mostly absent in this selection.
% The second dataset (\emph{Starting events}) is based on a geometrical selection and contains only muon tracks that start in the detector volume and have a minimal track length of $400$ m. The last dataset
The other two datasets are based on a geometrical Monte Carlo-based selection. One contains muon tracks starting outside the detector volume with a minimal track length of $700$~m (\emph{Through-going events}), the other muon tracks that start in the detector volume and have a minimal track length of $400$~m (\emph{Starting events}).
% One contains only muon tracks that start in the detector volume and have a minimal track length of $400$~m (\emph{Starting events}), the other only muon tracks starting outside the detector volume with a minimal track length of $700$~m (\emph{Through-going events}).

\subsection{Angular resolution}
\labsec{angres}

To illustrate the performance of the new reconstruction,  \reffig{ssreco_datasets} shows the median angular difference between true muon direction and reconstructed muon direction for the three classes of track-like events and the three likelihood formulations. {These are} compared to  \splinereco default settings, i.e. without any of the {modifications} (see \refsec{oldrecos} for a description of these {modifications}). It can be seen that \ssreco yields up to 20\% better angular resolutions at high energies for through-going tracks, and up to a factor 2 better resolutions for starting tracks. The likelihood that looks only at the first hit ($L_{\mathrm{1st}}$) performs generally as good or better than the other two, which is the known behavior that is observed for the prevailing reconstructions (\refsec{oldrecos}) and understood as mitigation of the {unphysical} hypothesis and the uncertainties of the ice model. The similarity of all likelihoods for through-going tracks indicates that the stochastic modelling in \ssreco improves the overall data description and narrows the advantage of $L_{\mathrm{1st}}$, even though the slight difference in outcomes shows that {the modelling is not perfect.} The difference in outcomes is to be compared with the respective difference for \splinereco (compare \splinereco + $L$ and \splinereco + $L_{\mathrm{1st}}$ in \reffig{pandel_splinereco}), which is much larger. {For starting tracks, a clear advantage for $L_{\mathrm{1st}}$ remains.} A potential explanation for this behavior is that the seed reconstruction used for these tracks is {often skewed} and the pure energy fit to determine a fixed energy loss profile {subsequently} gives a result that is rather far from the true energy loss profile. In these cases, $L_{\mathrm{1st}}$ {best manages the resulting inaccuracies on the model.}
% {In these cases is again L$_{\mathrm{1st}}$ the one can best get by with all the resulting inaccuracies on the model.} 
% In such situations again likelihood (c) can best cope with the resulting mis-specificiation of the overall model.

\begin{figure*}[t]
    \centering
    \includegraphics[width=\textwidth]{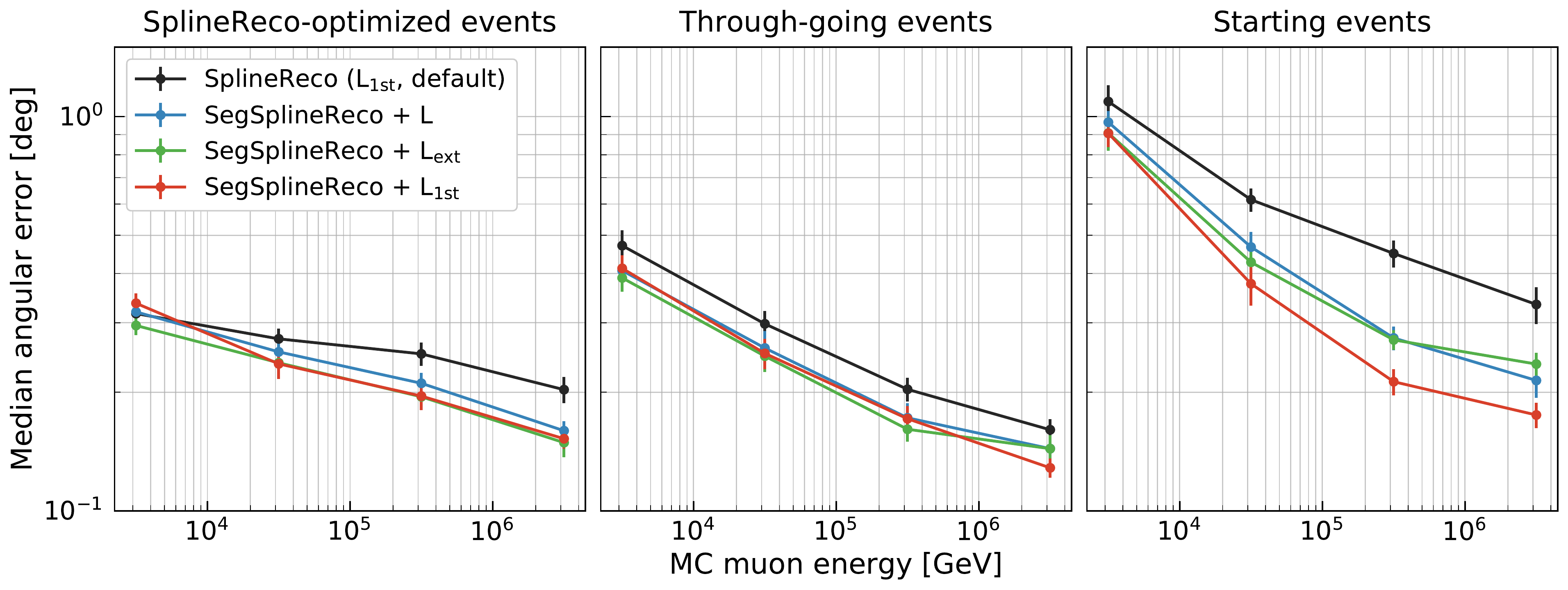}
    \caption{Median angular resolution as a function of MC muon energy calculated at the {interaction vertex} for three IceCube all-sky simulations: \emph{SplineReco-optimized} events, through-going and starting tracks. The \ssreco reconstruction is compared to \splinereco (black line). The three different likelihood models for \ssreco are compared: the standard unbinned likelihood $L$ (blue line), the extended unbinned likelihood $L_{\mathrm{ext}}$ (green line) and the unbinned likelihood for the first hit per DOM $L_{\mathrm{1st}}$ (red line). The statistical error on the median is calculated using bootstrapping.}
    \labfig{ssreco_datasets}
\end{figure*}

In \reffig{ssreco_datasets}, the initial energy loss pattern is obtained by performing a pure energy fit with \ssreco. This energy profile is then used as seed for the track reconstruction. \ssreco can also perform a simultaneous fit of track parameters and energy cascades. 
However, this second option gives worse angular resolution, as shown for $\mathrm{L}_{\mathrm{ext}}$ with joint vertex and energy optimization in \reffig{enfit_postjitter}. The worse performance is probably due to numerical instability issues of the high-dimensional problem. For this reason, by default the energies are always determined independently before the vertex parameters are optimized.

\reffig{enfit_postjitter} also shows how \ssreco with $L_{\mathrm{1st}}$ compares to \splinereco with (``max'') and without {modifications} (``default").
One of those {modifications} is an energy-dependent convolution of the time \pdf of the first hit (post-jitter), which models absolute time detection uncertainty between DOMs. Since \ssreco improves the angular resolution only at the highest energy when compared to \splinereco with \emph{max settings}, this convolution has also been applied to the new reconstruction. As shown in the figure, besides an energy-dependent convolution, also a fixed-time resolution convolution has been implemented. However, the energy-dependent post-jitter convolution improves the resolution at all energies. Therefore, the energy-dependent convolution is used as default setting in \ssreco. If no energy estimator is available, a $4.5$~ns convolution gives almost comparable results. 
{Other modifications are not really applicable to \ssreco}, like the effective stochastic loss profile, since they are {already naturally captured in the explicit stochastic modelling within \ssreco.} It can be seen in \reffig{ssreco_datasets_postjitter} that standard \ssreco is on par or slightly better than \splinereco with these settings. The extra energy-dependent time \pdf convolution is further improving the resolution by a few percent in all datasets. 

\begin{figure*}[htpb]
    \centering
    \includegraphics[width=0.7\textwidth]{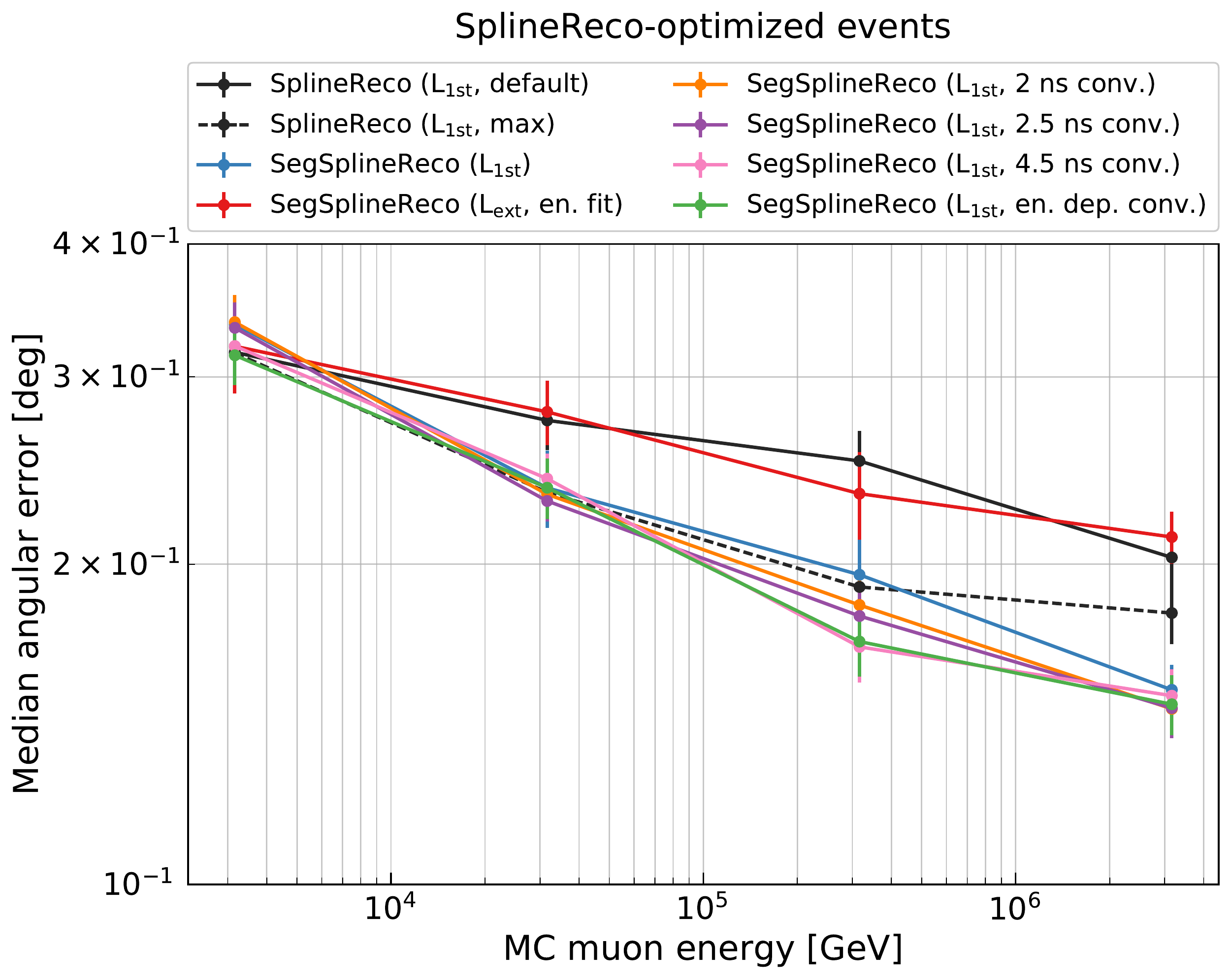}
    \caption{Median angular resolution as a function of MC muon energy {calculated at the interaction vertex} for the \emph{SplineReco-optimized} events. The \splinereco with default (full line) and max settings (dashed line) are shown in black. The standard \ssreco (blue line) is compared with \ssreco ($ \mathrm{L}_{\mathrm{ext}}$) with jointly fitted energy cascades (red line). The \ssreco reconstruction using $L_{\mathrm{1st}}$ with energy-dependent \pdf convolution is compared with a fixed convolution of 2~ns, 2.5~ns and 4.5~ns.}
    \labfig{enfit_postjitter}
\end{figure*}
\begin{figure*}[htpb]
    \centering
    \includegraphics[width=\textwidth]{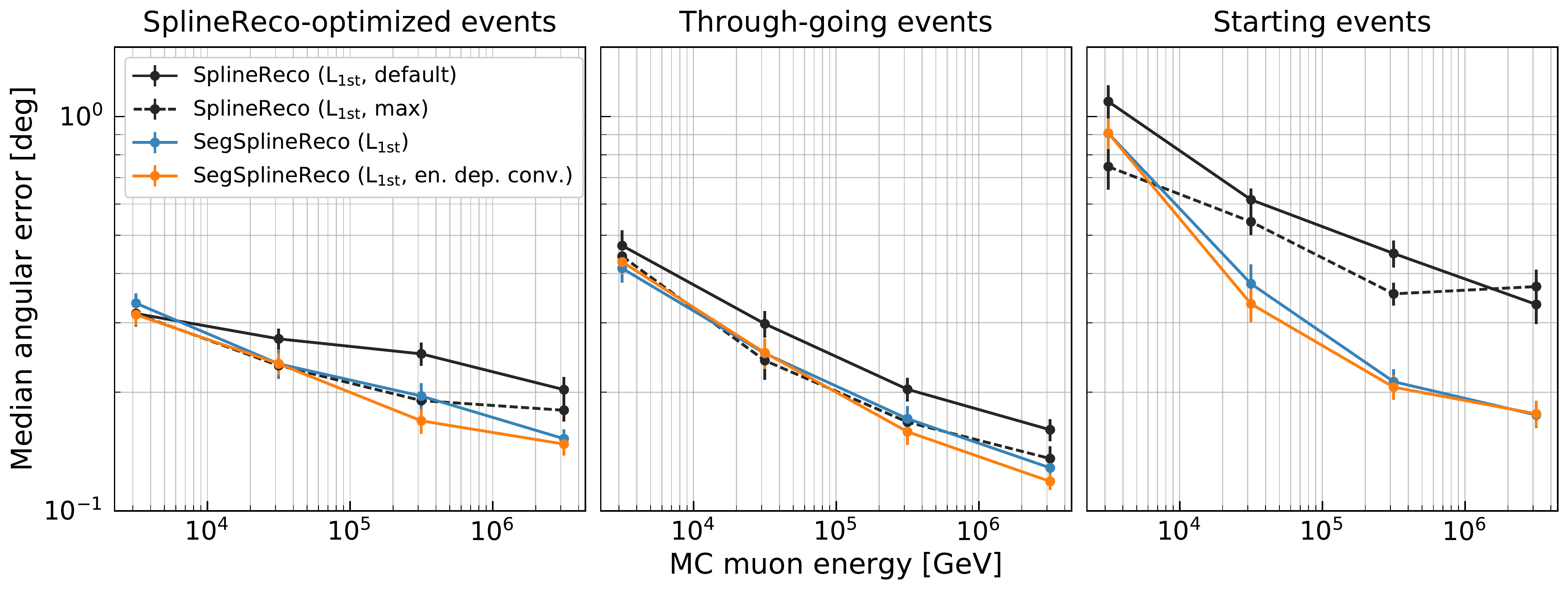}
    \caption{Median angular resolution as a function of MC muon energy {calculated at the neutrino interaction point} for three IceCube all-sky simulations: \emph{SplineReco-optimized} events, through-going and starting tracks. The \ssreco reconstruction using the $L_\mathrm{1st}$ likelihood with (orange line) and without (blue line) energy-dependent convolution (post-jitter) is compared to \splinereco with default settings (solid black line) and max settings (dashed black line). The statistical error on the median is calculated using bootstrapping.}
    \labfig{ssreco_datasets_postjitter}
\end{figure*}

\subsection{Angular error estimation}
\labsec{sec:angular_error_estimation_new}
As discussed in \refsec{subsec:ssreco_error}, two methods for the calculation of uncertainty contours are implemented in the new reconstruction. \reffig{contour_comparison} shows the uncertainty contours for two example events. {Additionally, the marginalized \pdf from the MCMC samples is indicated assuming a flat {prior}.} In general, the uncertainty contour from the analytic Hessian (\emph{Method 1}) is smaller than the paraboloid fit (\emph{Method 2}) for all events that show some non-Gaussian behavior. 
The quality of the uncertainty estimation can be judged by its coverage. 

\begin{figure*}[t]
    \centering
    \subfloat[Event with Gaussian minimum.]{
	{ \includegraphics[width=0.45\textwidth]{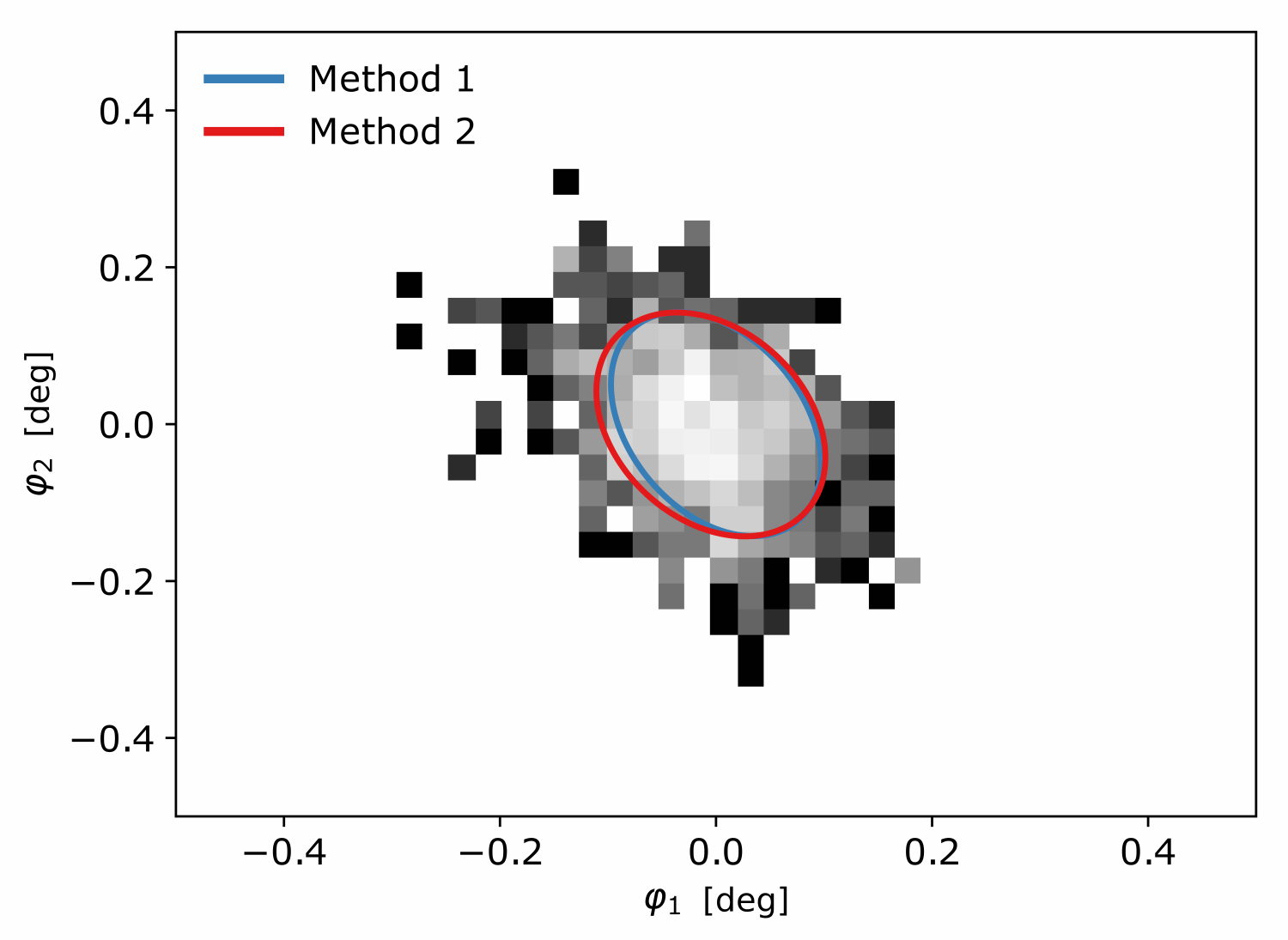} }
    }
   \subfloat[Event with slightly non-Gaussian minimum.]{
	{ \includegraphics[width=0.45\textwidth]{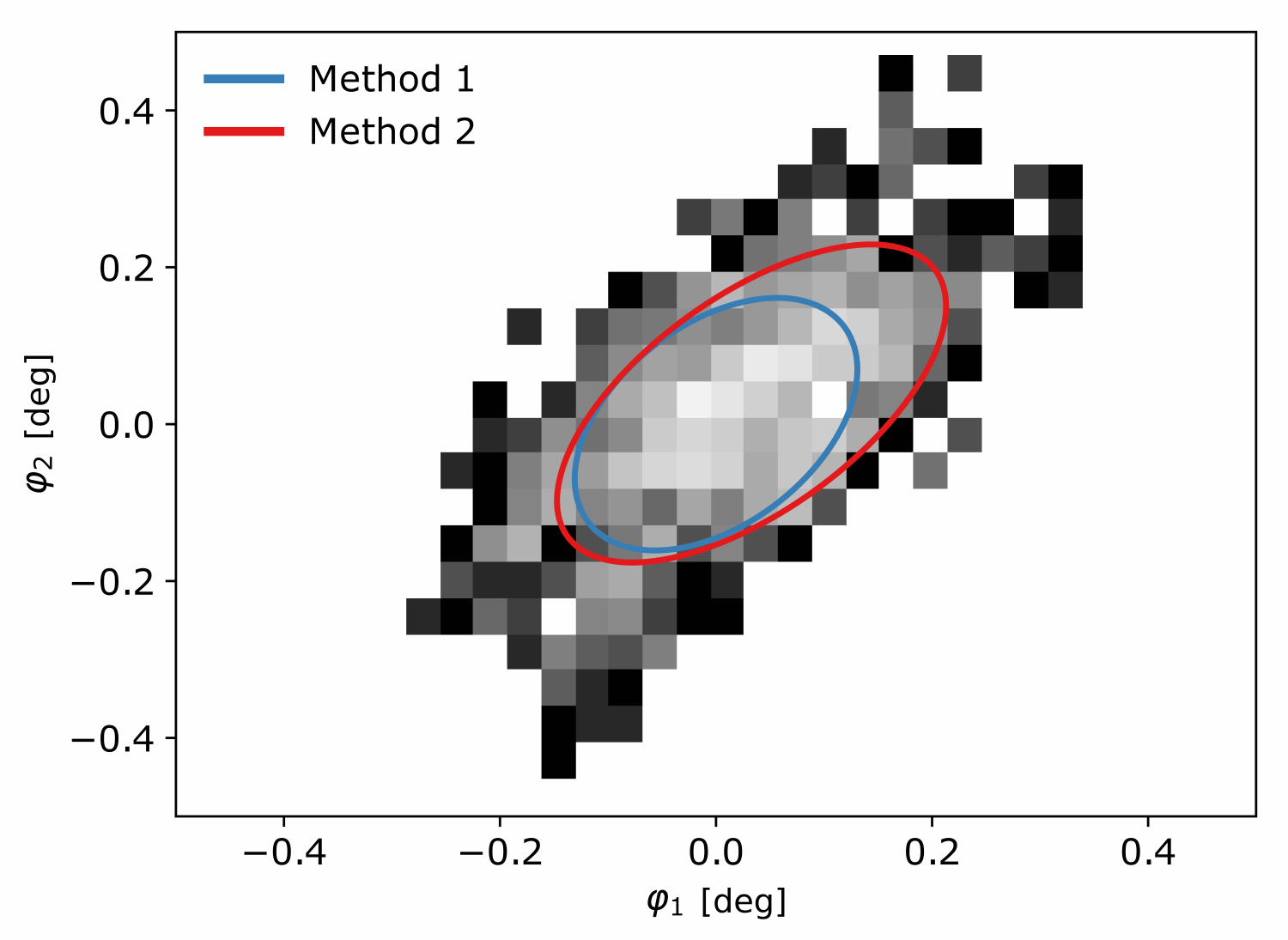} }
	}
    \caption{Comparison of both uncertainty  calculations  for  two  example  events. The two $68\%$ uncertainty ellipses are shown on top of the MCMC samples, which indicate the true marginalized density assuming a flat prior.}
    \labfig{contour_comparison}
\end{figure*} 
\begin{figure*}[ht]
    \centering
    \includegraphics[width=\textwidth]{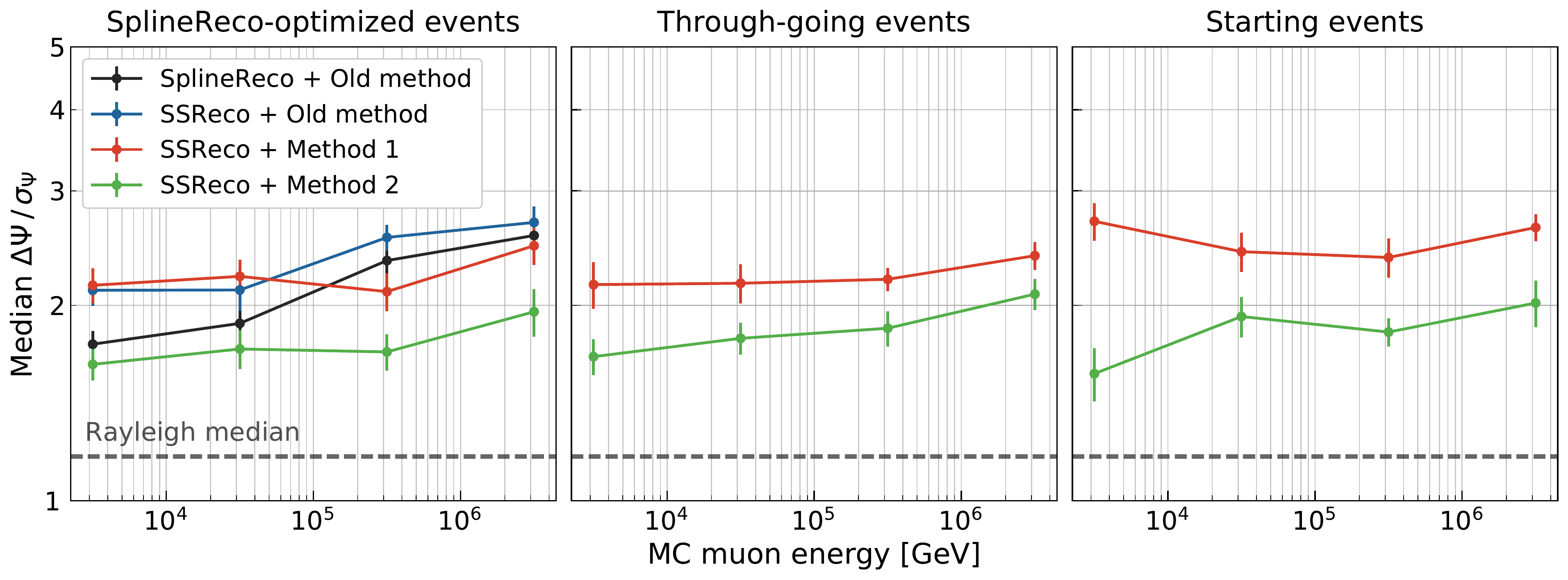} 
    \caption{Comparison of different angular error estimators for the \ssreco reconstruction with likelihood model (c) of the \emph{SplineReco-optimized}, \emph{through-going} and \emph{starting} events. The median of the pull ($\Delta \Psi / \sigma_{\Psi}$) is shown as a function of the MC muon energy. The statistical error on the median is calculated using bootstrapping. The median angular resolution $\Delta \Psi$ is obtained using the $L_\mathrm{1st}$ likelihood with energy-dependent post-jitter. $\sigma_{\Psi}$ has been calculated for the same muon track obtained from likelihood $L_\mathrm{1st}$ without energy-dependent post-jitter. The dashed gray line shows the ideal case of a Rayleigh distribution with $\sigma =1$ and median at 1.17. {The MC muon energy is calculated at the neutrino interaction point.}}
    \labfig{ssreco_datasets_error}
\end{figure*}

For simplicity we average major and minor axes of the uncertainty contour to create an averaged angular error for $\Delta\Psi$, the difference between reconstructed direction and true direction. This averaged uncertainty is calculated {using} $\sigma_{\Psi}=\sqrt{\frac{\sigma_1^2+\sigma_2^2}{2}}$. The {quantity} $\Delta \Psi/\sigma_{\Psi}$ is called the pull. Ideally, it is distributed like a Rayleigh distribution with $\sigma=1$ {which has a median of} $1.17$. \reffig{ssreco_datasets_error} shows the median pull for the three types of events. It compares both uncertainty calculations, and also includes the old 2-D paraboloid fit \cite{paraboloid} for the \emph{SplineReco-optimized} event class. It can be seen that the pull is generally flatter with the new uncertainty estimation, which shows that the shortcomings of fixed grid size and instability from profiling in the old uncertainty estimation are gone. However, a slight energy dependence remains, which {seems} to increase at the highest energy. This is reflected in the overall offset between the two curves in the median pull.
Overall, the median is larger than $1.17$ in any approach{,} which shows that the likelihood contours are underestimating the true uncertainty. {The underestimation partly stems from the fact that {the energies of the stochastic losses are fixed.} Additional contributions likely come from remaining model mis-specifications, like the assumption that all energy losses are modelled as point-like electromagnetic cascades at fixed distances along the track and from non-Gaussian behavior around the local optima.}

\subsection{Runtime}
\labsec{runtime}
 
\autoref{tab:running_times} shows a comparison of the running time for \splinereco and \ssreco with the corresponding uncertainty estimations for different energy ranges. The reconstruction times are the average {CPU} time for 100 standard likelihood evaluations, expressed in minutes. Each reconstruction has been performed on the \emph{SplineReco-quality} dataset. The new reconstruction is significantly more time consuming. 

On average, the running time per event for \ssreco that includes a standard vertex fit and the error estimation with \emph{Method 1} is of $\sim 1.25$~minutes, about 6 times larger than the running time required to perform a \splinereco vertex fit and uncertainty estimation with the \emph{traditional method}.  
% On average the running time per event for \ssreco that includes all evaluations of the energy fit, a standard vertex fit and the error estimation with \emph{Method 1} is about 6 times larger than \splinereco.
This run time fraction doubles when also the energy fit is performed for \ssreco and can get up to 100 times larger when performing the energy-dependent post-jitter convolution and using \emph{Method 2} for the uncertainty estimation.

\begin{table*}[h]
\caption{Running time comparison in minutes of some relevant reconstructions and uncertainty estimations. The reconstruction times are the average {CPU} time for 100 standard likelihood evaluations to give roughly comparable times that are independent of optimization algorithms. For \ssreco the distance between two modelled energy losses is 10~m.}
\resizebox{\textwidth}{!}{%
% \begin{tabular}{r| c|c|c|c|c|c|c}
\begin{tabular}{r c c ccccc}
\toprule[1pt]
& \multicolumn{2}{c}{\textsc{SplineReco}}  & \multicolumn{5}{c}{\textsc{SegmentedSplineReco}}   \\
% \cmidrule{2-8}
\cmidrule[0.6pt](lr){2-8} 
% \cmidrule[0.6pt](lr){4-8}
& \textsc{Reconstruction}  & \textsc{Uncertainty Estimation} & \multicolumn{3}{c}{\textsc{Reconstruction}}  & \multicolumn{2}{c}{\textsc{Uncertainty Estimation}} \\
\cmidrule(lr){2-8} 
% \cmidrule(lr){2-2} 
% \cmidrule(lr){3-3}
% \cmidrule(lr){4-6}
% \cmidrule(lr){7-8}
& Vertex fit ($L_{1\mathrm{st}}$, max settings) & \multirow{2}{*}{{Traditional Method}} & {Energy fit} &  {Vertex fit} & {With post-jitter} &  \multirow{2}{*}{{Method} 1} &  \multirow{2}{*}{{Method} 2} \\ 

& (100 evals) &  & (full) & (100 evals) & (100 evals) & &  \\ 
 
\midrule

1 TeV - 10 TeV  &  0.002 & 0.06  & 0.70 & 0.28 & 2.49 &0.09 &4.41\\ \hline
10 TeV - 100 TeV &   0.005  & 0.12  & 0.98 &  0.58 &  5.25 & 0.19 & 9.09\\ \hline
100 TeV - 1 PeV   & 0.010  &  0.22 & 1.46 &  1.20  & 10.29 &  0.32 & 16.38\\ \hline
1 PeV - 10 PeV & 0.017  &  0.34 & 2.22 & 1.88 & 19.32& 0.47 & 24.85 \\    

\bottomrule[1pt]
\end{tabular}}

\label{tab:running_times}
\end{table*}

\section{Discussion and outlook}
\labsec{disc}

We have introduced a new directional reconstruction for muons that explicitly models the stochastic losses in the likelihood by equidistant electromagnetic showers along the track. {For all track topologies and energies, the new reconstructions shows a better muon angular resolution.} %For all track topologies and energies, it improves the angular resolution. 
The improvement increases with energy as the stochastic modelling of the muon becomes more important. For throughgoing tracks the improvement is up to $10-20 \%$ at PeV energies. For starting tracks, it is much more pronounced and larger than a factor of $2$ above $100$ TeV. In addition to the reconstruction, the uncertainty estimation has been improved by a more numerically stable determination of the Hessian at the optimum. The resulting pull distribution is now nearly flat with energy using a likelihood that only looks at the first hit, which is a common procedure to mitigate systematics.
While the new reconstruction is more precise it is also more time consuming. Depending on the settings, the reconstruction including uncertainty estimation takes about 6 ({using} no energy-dependent convolution and Hessian Method 1) to over 100 times more processing time ({using} energy-dependent convolution and Hessian Method 2) per event than the previously state-of-the-art muon reconstruction \splinereco . {In later stages in typical event selection chains these processing times can still be feasible.} A few limiting factors remain. Currently the stochastic losses are modeled by pointlike electromagentic showers at fixed distances from each other. In reality, these losses have longitudinal emission profiles and are not equi-distant. In particular if the track passes close to a DOM such an unphysical hypothesis can bias the reconstruction. However, it is hard to see how such an assumption can be made more realistic in the parametric likelihood approach. A remaining drawback that can potentially be fixed is the currently neglected correlation between the true muon energy and the energy losses. Fitting simultaneously for the muon energy {could} stabilize and improve the energy loss solution. We leave that for future work.


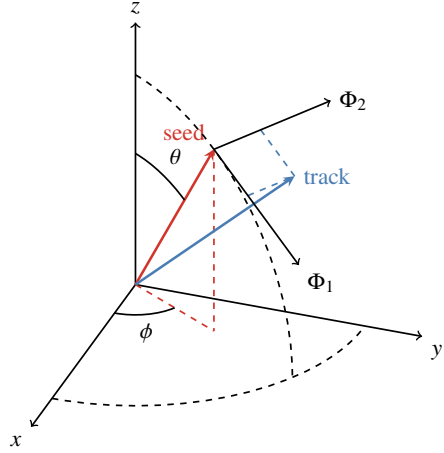
\begin{figure}[t]
\begin{center}
% \resizebox{0.7\columnwidth}{!}{
\resizebox{0.4\columnwidth}{!}{
\begin{tikzpicture}[scale=5,tdplot_main_coords, thick]
\coordinate (O) at (0,0,0);
\draw[thick,->] (0,0,0) -- (1,0,0) node[anchor=north east]{$x$};
\draw[thick,->] (0,0,0) -- (0,1,0) node[anchor=north west]{$y$};
\draw[thick,->] (0,0,0) -- (0,0,1) node[anchor=south]{$z$};
\tdplotsetcoord{P}{\rvec}{\thetavec}{\phivec}
\draw[-stealth,color=newred,very thick] (O) -- (P) node[above left] {seed};
\draw[dashed, color=newred] (O) -- (Pxy);
\draw[dashed, color=newred] (P) -- (Pxy);
\tdplotdrawarc{(O)}{0.2}{0}{\phivec}{anchor=north}{$\phi$}
%\tdplotsetthetaplanecoords{\phivec}
%\tdplotdrawarc[tdplot_rotated_coords]{(0,0,0)}{0.5}{0}%
%    {\thetavec}{anchor=south west}{$\theta$}

%set the rotated coordinate system so the x'-y' plane lies within the
%"theta plane" of the main coordinate system
%syntax: \tdplotsetthetaplanecoords{\phi}
\tdplotsetthetaplanecoords{\phivec}

%draw theta arc and label, using rotated coordinate system
\tdplotdrawarc[tdplot_rotated_coords]{(0.0,0.0,0)}{0.5}{0}{\thetavec}{anchor=south west}{$\theta$}

%draw some dashed arcs, demonstrating direct arc drawing
\draw[dashed,tdplot_rotated_coords] (\rvec,0,0) arc (0:90:\rvec);
\draw[dashed] (\rvec,0,0) arc (0:90:\rvec);

%translate the rotated coordinate system
%syntax: \tdplotsetrotatedcoordsorigin{point}

%\draw[-stealth,color=newblue,tdplot_rotated_coords] (0,0,0) -- (Px,Py,Pz);

%set the rotated coordinate definition within display using a translation
%coordinate and Euler angles in the "z(\alpha)y(\beta)z(\gamma)" euler rotation convention
%syntax: \tdplotsetrotatedcoords{\alpha}{\beta}{\gamma}
\tdplotsetrotatedcoords{\phivec}{\thetavec}{0}

\tdplotsetrotatedcoordsorigin{(P)}
\draw[-stealth,color=newblue,tdplot_rotated_coords, very thick] (0,0,-0.8) -- (.2,.2,0)node[anchor=west,color=newblue]{track};
\draw[dashed,color=newblue,tdplot_rotated_coords] (0.2,0,0) -- (.2,.2,0);
\draw[dashed,color=newblue,tdplot_rotated_coords] (0,.2,0) -- (.2,.2,0.0);

%use the tdplot_rotated_coords style to work in the rotated, translated coordinate frame
\draw[thick,tdplot_rotated_coords,->] (0,0,0) -- (.5,0,0) node[anchor=north west]{$\Phi_1$};
\draw[thick,tdplot_rotated_coords,->] (0,0,0) -- (0,.5,0) node[anchor=west]{$\Phi_2$};
%\draw[thick,tdplot_rotated_coords,->] (0,0,0) -- (0,0,.5) node[anchor=south]{$z'$};

\end{tikzpicture}
}
\end{center}
\caption[]{The auxiliary angular coordinates behave as Euclidean coordinates $\Phi_1$ and $\Phi_2$ in a unique tangential plane defined by the seed track.}
\labfig{fig_appendix_para1}
\end{figure}
\appendix
\section{Parametrizations of track orientations}
\labapp{appendix_angle_parametrizations}

Instead of zenith $\theta$ and azimuth $\phi$, it is more stable to change the parametrization of the angles to auxiliary coordinates that are independent of the position on the sphere. This is the case for both the optimization and uncertainty estimation procedures. Two common re-parametrizations are described in the following.

\subsection{Tangent plane parametrization}

At the current seed direction the tangent plane defines a new coordinate system which replaces $\theta$ and $\phi$ (see \reffig{fig_appendix_para1}). The coordinate axes are uniquely defined by a choice of orthogonal axes with respect to the seed track $\vec{s}=(s_x,s_y,s_z)$. An example choice is $\vec{e}_{\Phi_1}=(0, \frac{s_x}{n}, -\frac{s_y}{n})$, and $\vec{e}_{\Phi_2}=(-n, \frac{s_x \cdot s_y}{n} , \frac{s_x \cdot s_z}{n})$ with $n=\sqrt{{s_y}^2 + {s_z}^2}$. The current track hypothesis is located at $(\Phi_1,\Phi_2)$, where $\Phi_1$ and $\Phi_2$ are new optimization parameters. A drawback of this scheme is that for large deviations from the seed the parameters {lose} the meaning of an angle. This can be undesired behavior if uncertainty estimation is performed in these coordinates. The tangent plane parametrization is often used by the prevailing reconstructions.

\subsection{Rotation to (1,0,0)}
\labapp{app_equator_parametrization}

This parametrization is defined by the unique rotation $R$ of the current seed position to the $x$-axis (see \reffig{fig_appendix_para2}). The auxiliary angles $\varphi_1$ and $\varphi_2$ measure the rotated track coordinates relative to the rotated seed which aligns with the $x$-axis. The inverse rotation $R^{-1}$ applied to the current rotated track yields back the track orientation in the original coordinate system. An advantage of this parametrization is that $\varphi_1$ and $\varphi_2$ are always defined as angles and it is therefore suited for uncertainty contours that can measure tens of degrees in diameter. The parametrization is used by the prevailing uncertainty estimation precedure \cite{paraboloid}. It is also used by the new reconstruction \ssreco and the subsequent uncertainty estimation.

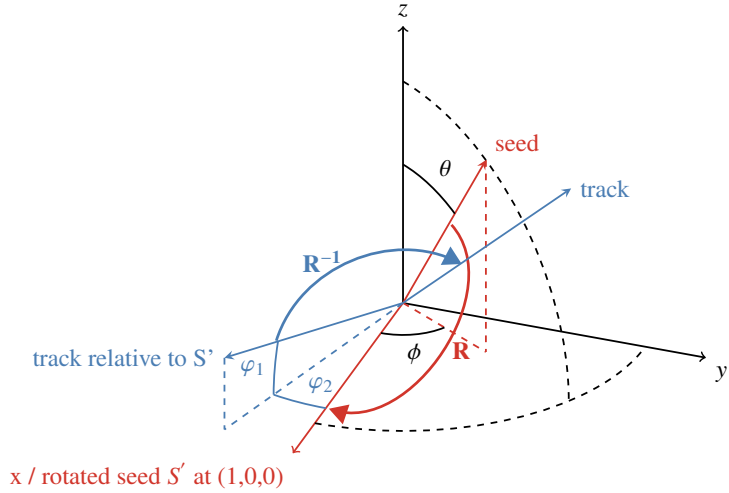
\begin{figure}[t]
\begin{center}
% \resizebox{0.99\columnwidth}{!}{
\resizebox{0.65\columnwidth}{!}{
\begin{tikzpicture}[scale=5,tdplot_main_coords, thick]
\coordinate (O) at (0,0,0);
\draw[thick,->,color=newred] (0,0,0) -- (1,0,0) node[anchor=north east, color=newred]{x / rotated seed $S^{'}$ at (1,0,0)};
\draw[thick,->] (0,0,0) -- (0,1,0) node[anchor=north west]{$y$};
\draw[thick,->] (0,0,0) -- (0,0,1) node[anchor=south]{$z$};
\tdplotsetcoord{P}{\rvec}{\thetavec}{\phivec}
\tdplotsetcoord{T}{1.0}{75}{-15}
\tdplotsetcoord{N}{1.0}{90}{0}
\draw[-stealth,color=newred] (O) -- (P) node[above right] {seed};
\draw[dashed, color=newred] (O) -- (Pxy);
\draw[dashed, color=newred] (P) -- (Pxy);
\tdplotdrawarc{(O)}{0.2}{0}{\phivec}{anchor=north}{$\phi$}
%\tdplotsetthetaplanecoords{\phivec}
%\tdplotdrawarc[tdplot_rotated_coords]{(0,0,0)}{0.5}{0}%
%    {\thetavec}{anchor=south west}{$\theta$}
    
%\coordinate (T) at (1.0,0.3,0.3);    
\draw[-stealth,color=newblue] (O) -- (T)node[anchor=east,color=newblue]{track relative to S'};
\draw[dashed, color=newblue] (O) -- (Txy);
\draw[dashed, color=newblue] (T) -- (Txy);
%set the rotated coordinate system so the x'-y' plane lies within the
%"theta plane" of the main coordinate system
%syntax: \tdplotsetthetaplanecoords{\phi}
\tdplotsetthetaplanecoords{\phivec}

% \draw [->,gray] (P.north) to [out=150,in=30] (1,0,0);
\draw[->,-triangle 60, color=newred, very thick](0,\rvec/5,\rvec/2.55) to[bend right=-75]node[ right]{$\mathbf{{R}}$}(0.7,0.01,0);

% \draw[->,-triangle 60, color=black, very thick](0.6,0,0) to[bend right=-70]node[left]{$\mathbf{{R}^{-1}}$} (0.15,\rvec/5,\rvec/2.55);

\draw[->,-triangle 60, color=newblue, very thick](0.6,-0.2,0.15) to[bend right=-50]node[left]{$\mathbf{{R}^{-1}}$} (0.2,\rvec/3,\rvec/2.65);

%draw theta arc and label, using rotated coordinate system
\tdplotdrawarc[tdplot_rotated_coords]{(0.0,0.0,0)}{0.5}{0}{\thetavec}{anchor=south west}{$\theta$}

%draw some dashed arcs, demonstrating direct arc drawing
\draw[dashed,tdplot_rotated_coords] (\rvec,0,0) arc (0:90:\rvec);
\draw[dashed] (\rvec,0,0) arc (0:90:\rvec);

\tdplotsetrotatedcoordsorigin{(P)}
\tdplotsetrotatedcoords{\phivec}{\thetavec}{0}
\draw[-stealth,color=newblue,tdplot_rotated_coords] (0.0,0.0,-0.8) -- (.2,.2,0)node[anchor=west,color=newblue]{track};

%translate the rotated coordinate system
%syntax: \tdplotsetrotatedcoordsorigin{point}

%set the rotated coordinate definition within display using a translation
%coordinate and Euler angles in the "z(\alpha)y(\beta)z(\gamma)" euler rotation convention
%syntax: \tdplotsetrotatedcoords{\alpha}{\beta}{\gamma}
%\tdplotsetrotatedcoords{\phivec}{\thetavec}{0}
\tdplotsetrotatedcoords{0}{-90}{0}
\tdplotsetrotatedcoordsorigin{(N)}

%\draw[thick,tdplot_rotated_coords,->] (0,0,0) -- (.5,0,0) node[anchor=north west]{};
%\draw[thick,tdplot_rotated_coords,->] (0,0,0) -- (0,.5,0) node[anchor=west]{};

\tdplotsetthetaplanecoords{-15}
\tdplotsetrotatedcoordsorigin{(O)}

\tdplotdrawarc[color=newblue]{(O)}{0.7}{0.0}{-15}{anchor=south west}{$\varphi_2$}
\tdplotdrawarc[tdplot_rotated_coords, color=newblue]{(0.0,0.0,0)}{0.7}{75}{90}{anchor=east}{$\varphi_1$}

%\draw[-stealth,color=newblue,tdplot_rotated_coords] (0,0,0) -- (.2,.2,0);
%\draw[dashed,color=newblue,tdplot_rotated_coords] (0.2,0,0) -- (.2,.2,0);
%\draw[dashed,color=newblue,tdplot_rotated_coords] (0,.2,0) -- (.2,.2,0.0);

%use the tdplot_rotated_coords style to work in the rotated, translated coordinate frame

%\draw[thick,tdplot_rotated_coords,->] (0,0,0) -- (0,0,.5) node[anchor=south]{$z'$};

\end{tikzpicture}
}
\end{center}
\caption[]{The auxiliary angular coordinates $\varphi_1$ and $\varphi_2$ are measured with respect to the rotated Seed track $S'$ that aligns with the x-axis. They behave as standard angular coordinates. }
\labfig{fig_appendix_para2}
\end{figure}

\section{Technical details of the uncertainty estimation}
\labapp{uncertainty_details}
This section describes the technical details of the uncertainty estimations introduced in section \refsec{subsec:ssreco_error}. From Wilks' theorem \cite{wilks} it follows that the quantity $\lambda=2 \cdot (\mathrm{logL}(\theta)-\mathrm{logL}(\theta_0))$ is approximately $\chi^2_n$-distributed with $n$ degrees of freedom if $n$ is the number of nested parameters. At the same time, the $\chi^2_n$-distribution is equivalent to the logarithmic difference of the n-dimensional Multivariate Gaussian distribution at its maximum with any other point drawn from the Gaussian \cite{chi2_fact}. Therefore, $\lambda$ can be approximated around $\theta_0$ with
\begin{align}
\lambda&=-2 \cdot (\mathrm{logL}(\theta)-\mathrm{logL}(\theta_0)) \\ &\approx (\theta-\theta_0) \cdot \mathrm{Cov_n}^{-1} \cdot (\theta-\theta_0) \\ &= (\theta-\theta_0) \cdot H_0 \cdot (\theta-\theta_0) \label{eq:h_0}
\end{align}
where $\mathrm{Cov_n}^{-1}$ is the inverse covariance matrix of the related n-dimensional Gaussian, which simultaneously is equivalent to {the Hessian matrix} $H_0$ of the log-likelihood function at the optimum $\theta_0$. We can therefore write for $-\mathrm{logL}(\theta)$ close to the optimum $\theta_0$
\begin{align}
   -\mathrm{log}L(\theta)&\approx (\theta-\theta_0) \cdot \frac{H_0}{2} \cdot(\theta-\theta_0) - \mathrm{log}L(\theta_0) \\ &\equiv
   (\theta-A) \cdot \frac{B}{2} \cdot(\theta-A) +C \\
   &=f(\theta;A,B,C) 
   \label{eq:parabola}
\end{align}
which is the formula for a general elliptic paraboloid in $n$ dimensions with position A, parameter matrix B/2 and offset C. With these relations it is clear that determination of $H_0$ allows {to determine} the covariance matrix, and thereby Gaussian contours, after matrix inversion. As discussed in \refsec{subsec:ssreco_error}, two methodologies to estimate $H_0$ have been developed and are described in the following. 
\paragraph{Method 1}
The first method calculates the Hessian matrix at the optimum, $H_0$, analytically. Advances in automatic differentiation have only recently made this feasible, in particular we used \emph{autograd}\footnote{\url{https://github.com/HIPS/autograd}} to crosscheck the implementation.
\paragraph{Method 2}
The second method fits a paraboloid to samples $\mathrm{logL}(\hat{\theta}_i)$ of the negative likelihood function near the optimum. The samples are obtained with an affine-invariant particle-based Markov-Chain sampler, \emph{emcee} \cite{Foreman_Mackey_2013}. The particle positions are initialized as samples drawn from a Gaussian distribution with covariance $H_0^{-1}$, where $H_0$ is the analytically calculated Hessian at the optimum (\emph{Method 1}). This initialization skips the burn-in phase completely if the log-likelihood optimum is nearly Gaussian, and otherwise drastically speeds up convergence. For the 6-dimensional sampling ($x$, $y$, $z$, $t$ and two angle dimensions) a burn-in phase of 2000 evaluations (100 particles with 20 iterations) followed by another 2000 samples is usually enough. In the second step a Levenberg-Marquardt algorithm is used to minimize the loss function 
\begin{align}
\mathrm{Loss}=0.5 \cdot \sum_i \rho\left([-\mathrm{logL}(\theta_i)-f(\theta_i;A,B,C)]^2\right)
\end{align}
and thereby fit a paraboloid shape to the samples. The term $\rho(x)=x$ yields the standard least-square loss. Empirically an often more robust fit is achieved with $\rho(x)=2 \cdot (\sqrt{1+x}-1)$, which represents a ``soft-l1" distance that better handles non-Gaussianities or irregularities in the likelihood samples.\footnote{{See the documentation of Levenberg-Marquardt algorithm on the scipy webpage for more details: \url{https://docs.scipy.org}.}}
The parameter $A$ represents an n-dimensional mean and $C$ represents a 1-dimensional offset. The $n \times n$ parameter matrix $B$ has to be positive definite. This is ensured by a parametrizion in terms of its lower-triangular Cholesky-decomposition, which involves $n$ strictly positive parameters on the diagonal and $\frac{n^2-n}{2}$ parameters for the lower-triangular off-diagonal elements. The total number of parameters of the least-square fit {is} then $\frac{n^2+n}{2}+n+1$. If the resulting mean $A$ is sufficiently different from $\theta_0$, or if the offset $C$ is sufficiently different from $-\mathrm{logL}(\theta_0)$, this can indicate strong non-Gaussian behavior that even a modified least-square fit can not handle. In general this method yields slightly wider and more conservative contours than the analytic calculation (see also \reffig{contour_comparison}), because non-Gaussianities in combination with the soft least-square fit widen the tails of the paraboloid solution. 

\acknowledgments
The IceCube collaboration acknowledges the significant contributions to this manuscript from Federica Bradascio and Thorsten Gl\"usenkamp. The authors gratefully acknowledge the support from the following agencies and institutions: USA {\textendash} U.S. National Science Foundation-Office of Polar Programs,
U.S. National Science Foundation-Physics Division,
U.S. National Science Foundation-EPSCoR,
Wisconsin Alumni Research Foundation,
Center for High Throughput Computing (CHTC) at the University of Wisconsin{\textendash}Madison,
Open Science Grid (OSG),
Extreme Science and Engineering Discovery Environment (XSEDE),
Frontera computing project at the Texas Advanced Computing Center,
U.S. Department of Energy-National Energy Research Scientific Computing Center,
Particle astrophysics research computing center at the University of Maryland,
Institute for Cyber-Enabled Research at Michigan State University,
and Astroparticle physics computational facility at Marquette University;
Belgium {\textendash} Funds for Scientific Research (FRS-FNRS and FWO),
FWO Odysseus and Big Science programmes,
and Belgian Federal Science Policy Office (Belspo);
Germany {\textendash} Bundesministerium f{\"u}r Bildung und Forschung (BMBF),
Deutsche Forschungsgemeinschaft (DFG),
Helmholtz Alliance for Astroparticle Physics (HAP),
Initiative and Networking Fund of the Helmholtz Association,
Deutsches Elektronen Synchrotron (DESY),
and High Performance Computing cluster of the RWTH Aachen;
Sweden {\textendash} Swedish Research Council,
Swedish Polar Research Secretariat,
Swedish National Infrastructure for Computing (SNIC),
and Knut and Alice Wallenberg Foundation;
Australia {\textendash} Australian Research Council;
Canada {\textendash} Natural Sciences and Engineering Research Council of Canada,
Calcul Qu{\'e}bec, Compute Ontario, Canada Foundation for Innovation, WestGrid, and Compute Canada;
Denmark {\textendash} Villum Fonden and Carlsberg Foundation;
New Zealand {\textendash} Marsden Fund;
Japan {\textendash} Japan Society for Promotion of Science (JSPS)
and Institute for Global Prominent Research (IGPR) of Chiba University;
Korea {\textendash} National Research Foundation of Korea (NRF);
Switzerland {\textendash} Swiss National Science Foundation (SNSF);
United Kingdom {\textendash} Department of Physics, University of Oxford.

\bibliographystyle{JHEP_mod}
\bibliography{references}

\end{document}